%% file: main.tex
\documentclass[twocolumn]{aastex701}

\usepackage[
    mode=match,
    range-units=single,
    range-phrase=\text{--},
    retain-explicit-plus
]{siunitx}
\usepackage{amsmath}
\usepackage{caption}
\usepackage[dvipsnames]{xcolor}

\newcommand{\safeminus}{\texorpdfstring{$-$}{−}}
\DeclareSIUnit{\pc}{pc} 
\DeclareSIUnit{\kpc}{kpc} 
\DeclareSIUnit{\yr}{yr} 
\DeclareSIUnit{\Jy}{Jy} 

\begin{document}

\title{Probing the Interstellar Medium towards the Double Pulsar and the Gum Nebula using Low-Frequency Pulse Broadening Measurements}

\author[orcid=0000-0001-6840-4114, sname='Lee']{Christopher P. Lee}
\affiliation{International Centre for Radio Astronomy Research, Curtin University, Kent Street, Bentley, WA 6102, Australia}
\email[show]{christopher.lee@icrar.org}

\author[orcid=0000-0002-8383-5059, sname='Bhat']{N. D. Ramesh Bhat}
\affiliation{International Centre for Radio Astronomy Research, Curtin University, Kent Street, Bentley, WA 6102, Australia}
\email{ramesh.bhat@curtin.edu.au}

\author[orcid=0000-0001-8845-1225, sname='Meyers']{Bradley W. Meyers}
\affiliation{Australian SKA Regional Centre (AusSRC), Curtin University, Kent Street, Bentley, WA 6102, Australia}
\affiliation{International Centre for Radio Astronomy Research, Curtin University, Kent Street, Bentley, WA 6102, Australia}
\email{bradley.meyers@curtin.edu.au}

\author[orcid=0000-0003-2519-7375, sname='van Straten']{Willem van Straten}
\affiliation{Manly Astrophysics, 15/41-42 East Esplanade, Manly, NSW 2095, Australia}
\email{vanstraten.willem@gmail.com}

\begin{abstract}

\input{0-Abstract}

\end{abstract}


\input{1-Introduction}
\input{2-Data}
\input{3-Analysis-Results}
\input{4-Discussion}
\input{5-Conclusions}
\input{6-Acknowledgements}

\appendix

\input{7-Appendix}

\bibliography{Articles,Software,My-Lead-Author,My-Co-Author,Theses,Books}
\bibliographystyle{aasjournalv7}

\end{document}

%% file: 0-Abstract.tex
Long-term timing of the double pulsar system (PSR J0737$-$3039A/B) has enabled some of the most stringent constraints on strong-field gravity.
Yet, despite the system's high profile, there have been very few observations at frequencies below \qty{300}{\MHz}, where pulse broadening due to interstellar scattering becomes significant.
Measurements of the pulse broadening timescale and its frequency scaling, $\tau_\mathrm{s}\propto\nu^\alpha$ (where $\alpha$ is the scattering spectral index), provide valuable constraints on the structure of the scattering medium.
Here we report on simultaneous observations of J0737$-$3039A and five other nearby pulsars (J0630$-$3824, J0742$-$2822, J0820$-$4114, J0835$-$4510, J0837$-$4135) in the frequency range \qtyrange{109}{231}{\MHz} using the Murchison Widefield Array (MWA).
These observations have enabled the highest-quality detections ever published for these pulsars at low frequencies.
For J0737$-$3039A, J0742$-$2822, and J0837$-$4135, we measure $\alpha$ to be $-4.1(1)$, $-3.8(1)$, and $-4.1(1)$, respectively.
These measurements provide new constraints on the density and turbulence parameters of the Gum Nebula region, and we discuss implications for models of the Galactic electron density distribution.
Our current data yield a dispersion measure (DM) precision of $\sim$\qty{4e-4}{\per\cm\cubed\pc} for J0737$-$3039A, which can be further improved with the MWA Phase III to reach $\sim$\qty{1e-4}{\per\cm\cubed\pc}, far exceeding the precision of previous studies.
The prospects for precisely constraining temporal variations in the DM and pulse broadening timescale strengthen the importance of regular monitoring of J0737$-$3039A at low frequencies, with the eventual goal of contributing to long-term timing studies.

%% file: 1-Introduction.tex
\section{Introduction}\label{sec:dp:intro}

High-precision timing of millisecond pulsars (MSPs) in relativistic binary systems has proven time and again to be a powerful method for probing fundamental physics.
In particular, pulsar timing can be used to measure relativistic orbital dynamics, typically described using the ``post-Keplerian" (PK) timing model \citep{DD1985,DD1986}.
The measurement of two PK parameters can be used to determine the masses of the pulsar and its companion, which can be used to constrain the equation of state of ultra-dense matter \citep[e.g.][]{Rutherford2024,Koehn2025,Basu2025}.
If three or more PK parameters are measured, they can be used to perform self-consistency tests of general relativity or other theories of gravity \citep[e.g.][]{Freire2024,VKrishnan2025}.
The double pulsar system (PSR J0737$-$3039A/B) is the current gold standard for performing tests of gravity.
It is the only known binary system to contain two active radio pulsars: a mildly-recycled pulsar with a period of \qty{23}{\ms} and a non-recycled pulsar with a period of \qty{2.8}{\s} \citep[A and B, respectively;][]{Burgay2003,Lyne2004}.
These pulsars are in an ultra-relativistic edge-on orbit with a period of \qty{2.45}{\hour}.
Timing of the double pulsar has led to the measurement of six PK parameters and enabled the most precise test ever of the radiative properties of gravity \citep{Kramer2021PhRvX}.

Pulsar timing requires correcting for cold-plasma dispersion imparted by the ionized interstellar medium (IISM), which is characterized by a chromatic delay $t_\mathrm{DM}\propto\mathrm{DM}\times\nu^{-2}$ (where $\nu$ is the electromagnetic frequency).
The delay is proportional to the dispersion measure (DM), which is the integrated column density of free electrons along the line of sight.
Changes in the line of sight through the IISM can cause DM variations on timescales of months to years \citep[e.g.][]{Backer1993,Hobbs2004,You2007,Jones2017,Keith2024}.
Additionally, pulsars near the ecliptic exhibit annual variations due to the changing line of sight through the interplanetary medium \citep[e.g.][]{Tiburzi2021,Reardon2023}.
These variations need to be modeled accurately in order to maximize pulsar timing precision, which requires precise and regular measurements of the DM \citep[e.g.][]{Iraci2024,Iraci2025}.
As the observing campaign of the double pulsar continues and more high-quality data are obtained, the precision on PK parameters will be increasingly limited by the accuracy of the DM model.
This motivates the exploration of lower-frequency observations to potentially obtain better DM precision and to investigate other IISM propagation effects such as interstellar scattering.

Propagation through density variations in the IISM causes pulsar signals to be scattered over a range of angles \citep[see][]{Rickett1977,Rickett1990}.
The scattered signals arrive later due to the longer path lengths, causing the pulse profile to be broadened on the trailing side.
For simplicity, it is common to assume that the scattering medium can be approximated by a single thin screen of infinite transverse extent located approximately midway to the pulsar \citep[e.g.][]{Scheuer1968,Williamson1972}.
In this model, the scatter-broadened profile can be described by the convolution of the intrinsic profile with a single-sided decaying exponential with a time constant $\tau_\mathrm{s}$ (referred to as the scattering timescale or the pulse broadening timescale).
Additionally, interference between the scattered signals causes a pattern of frequency- and time-dependent intensity variations on the plane of the observer.
The relative motion of the pulsar, observer, and scattering medium causes the pattern to move over the observer, resulting in intensity scintillations in the observed signal (i.e.\ diffractive interstellar scintillation or DISS) with a decorrelation bandwidth $\Delta\nu_\mathrm{d}$ and timescale $\Delta t_\mathrm{d}$.
The condition for wave interference relates $\tau_\mathrm{s}$ and $\Delta\nu_\mathrm{d}$ as
\begin{equation}\label{eq:scat-scin-relation}
    2\pi\tau_\mathrm{s}\Delta\nu_\mathrm{d}=C_1,
\end{equation}
where $C_1\sim1$ is a constant that depends on the spectral and geometric properties of the scattering medium along line of sight \citep[][Equation 4.29]{LorimerKramer2012}.
The spectrum of $\tau_\mathrm{s}$ and $\Delta\nu_\mathrm{d}$ is described by a power law, $\tau_\mathrm{s}\propto1/\Delta\nu_\mathrm{d}\propto\nu^\alpha$, where $\alpha$ is the scattering spectral index \citep[e.g.][]{Cordes1986,Bhat2004}.
For a Kolmogorov turbulence spectrum, $\alpha=-4.4$ \citep[e.g.][]{LJ1976}.
However, if the scattering is dominated by a single characteristic scale, then geometric optics predicts $\alpha=-4$ \citep[sometimes referred to as a ``square-law" spectrum; e.g.][]{Cronyn1970,Lang1971}.
Measurements of $\alpha$ can be used to develop a physical interpretation of the scattering medium along a given line of sight, including its distribution and turbulence.
On average, observations of a large number of pulsars have found $\alpha\sim-4$ \citep[e.g.][]{Bhat2004,Geyer2017,Kirsten2019,Oswald2021,Jing2025}.

Characterizing the effects of scattering is important for quantifying the noise contribution to pulsar timing residuals \citep[e.g.][]{Coles2010,Cordes2010,Kramer2021PhRvX} and can provide insights into the physical properties and dynamics of foreground structures in the intervening IISM \citep[e.g.][]{Reardon2019,Askew2023,Reardon2025}.
The double pulsar is located in the Gum Nebula region, which is a large structure in the local IISM explicitly modeled in Galactic electron density models such as NE2025 \citep{NE2025}.
Measurements of $\tau_\mathrm{s}$ from pulsars in the vicinity of the Gum Nebula can be used to inform and validate these models.

Interstellar scintillation has been studied in detail for pulsar A the double-pulsar system (hereafter, we refer to pulsar A as either J0737$-$3039A or the double pulsar), resulting in constraints on the scattering medium, orbital geometry, and distance \citep{Coles2005,Rickett2014,Askew2026}.
However, pulse broadening has never been directly measured as it is significant only at low frequencies.
To date, there have been very few observations of the double pulsar below \qty{300}{\MHz}.
A recent study by \citet{Krishnakumar2025} observed the pulsar down to \qty{275}{\MHz} using the upgraded Giant Metrewave Radio Telescope (uGMRT), but did not detect any significant pulse broadening.
\citet{Kondratiev2016} reported a detection with the Low-Frequency Array (LOFAR) at \qty{150}{\MHz}; however, the sensitivity was greatly limited by the low declination of the source ($\mathrm{S}/\mathrm{N}\approx6$ in \qty{60}{\minute}).
The Murchison Widefield Array (MWA) has a much better vantage point, with the source transiting at a zenith angle of just \qty{4}{\degree}.
Given the short period and moderate DM (\qty{48.91}{\per\cm\cubed\pc}) of the double pulsar, the MWA frequency band is well suited to measure pulse broadening.
Indeed, previous detections with the MWA show visible signs of pulse broadening at \qty{154}{\MHz} and \qty{185}{\MHz} \citep{XueThesis,Bhat2023II,Lee2025}.

Motivated by these earlier detections, we have collected a set of dedicated observations of the double pulsar with the MWA between \qtyrange{109}{231}{\MHz}.
The aim of this work is to use these observations to probe the IISM in the direction of the double pulsar and the Gum Nebula, as well as to investigate the timing potential of the double pulsar at low frequencies.
In Section~\ref{sec:dp:data}, we described the observations, calibration, and data processing.
In Section~\ref{sec:dp:analysis}, we describe the analysis and present the DM and pulse broadening measurements.
In Section~\ref{sec:dp:discussion}, we discuss a joint analysis with higher-frequency scintillation data, how our data compare with past measurements and Galactic electron density models, and the prospects for monitoring the double pulsar at low frequencies.
In Section~\ref{sec:dp:conclusions}, we present our conclusions and discuss future work.

%% file: 2-Data.tex
\section{Data}\label{sec:dp:data}

\subsection{Observations}

The observations were collected using the MWA, a low-frequency aperture array telescope located at Inyarrimanha Ilgari Bundara, the CSIRO Murchison Radio-astronomy Observatory in Western Australia \citep{Tingay2013,Wayth2018}.
The MWA comprises 256 ``tiles" of 16 dual-polarized dipole antennas and operates in the frequency range \qtyrange{70}{300}{\MHz}.
The data were recorded using the MWAX Voltage Capture System \citep[VCS;][]{Morrison2023}, which records data for a \qty[number-unit-product=-]{30.72}{\MHz} instantaneous bandwidth divided into $24\times\qty[number-unit-product=\text{-}]{1.28}{\MHz}$ ``coarse" channels with a time resolution of \qty{781.25}{\ns}.
Any set of 24 coarse channels within the frequency range of the receivers can be selected for VCS recording.
The frequency setup can be optimized for a large fractional frequency span by distributing the channels in a non-contiguous or ``picket fence" configuration.
Picket fence observations have previously been used to obtain precise measurements of dispersion and Faraday rotation in pulsar observations with the MWA \citep{Xue2019,Kaur2019,Kaur2022}.
For this work, we chose to distribute the bandwidth into $4\times\qty[number-unit-product=\text{-}]{7.68}{\MHz}$ subbands (i.e.\ 6 coarse channels per subband) between $\sim$\qtyrange{109}{231}{\MHz}.
The subbands were arranged with log-uniform spacing to provide optimal constraints on the scattering spectral index.
The observing setup is summarized in Table~\ref{tab:obs-setup}.

\begin{deluxetable}{rl}
\tablewidth{0pt}
\tablecaption{Observation setup. \label{tab:obs-setup}}
\tablehead{Parameter [unit] & Value}
\startdata
Observation Length [\unit{\s}]     & 9000 \\
Subobservation Length\tablenotemark{a} [\unit{\s}]  & 3000 \\
Time Resolution [\unit{\us}]       & 100 \\
Total Bandwidth [\unit{\MHz}]      & 30.72 \\
Subband Bandwidth [\unit{\MHz}]    & 7.68 \\
Channel Bandwidth [\unit{\MHz}]    & 0.01 \\
Lower Frequencies [\unit{\MHz}]    & 109.44, 138.88, 176.00, 223.36 \\
Central Frequencies [\unit{\MHz}]  & 113.28, 142.72, 179.84, 227.20 \\
Upper Frequencies [\unit{\MHz}]    & 117.12, 146.56, 183.68, 231.04 \\
\enddata
\tablenotetext{a}{The primary beam was repointed between subobservations.}
\end{deluxetable}

\begin{deluxetable}{lcccc}
\tablewidth{0pt}
\tablecaption{Summary of observations. \label{tab:obs-summary}}
\tablehead{
    Label &
    \colhead{Date} &
    \colhead{$N_\mathrm{tile}$\tablenotemark{a}} &
    \colhead{$\tilde{B}_\mathrm{c}$\tablenotemark{b}} &
    \colhead{FWHM\tablenotemark{c}} \\[-5pt]
    &
    \colhead{(UTC)} &
    &
    \colhead{[\unit{\m}]} &
    \colhead{[\unit{\arcminute}]}
}
\startdata
Epoch A & 2023-12-20 & 104--105 &  227 & 19.63--41.44 \\
Epoch B & 2024-01-20 & 115--118 &  240 & 18.57--39.20 \\
Epoch C & 2025-01-11 &       85 & 3215 &  1.39--2.93 \\
Epoch D & 2025-02-24 & 109--111 & 3182 &  1.40--2.96 \\
\enddata
\tablenotetext{a}{Number of ``good" tiles (i.e.\ those with no known issues and a usable calibration solution).}
\tablenotetext{b}{Effective characteristic baseline length of the array \citep{Meyers2026}.}
\tablenotetext{c}{Full width at half maximum of the tied-array beam over the frequency range of the observation.}
\end{deluxetable}

During observing semesters 2023B and 2024B, we observed two full orbits of the double pulsar in observing sessions separated by $\sim$1 month.
The total timing baseline of the four observing sessions is $\sim$14 months.
Each observation was \qty{150}{\minute} in length ($\sim$1.02 orbits), comprising three contiguous \qty[number-unit-product=-]{50}{\minute} subobservations.
The tile pointing center, which is fixed in azimuth and elevation while recording data, was changed between subobservations to reduce the sensitivity loss due to beam attenuation in the direction of the double pulsar.
The four observations are summarized in Table~\ref{tab:obs-summary}, where we have labeled them chronologically as Epochs A--D (details about each subobservation are provided in Appendix~\ref{app:observations}).
For Epochs A and B, the observations were collected in the Phase II Compact configuration \citep{Wayth2018}, whereas for Epochs C and D, the observations were collected in extended array configurations using a combination of Phase I and Phase II tiles.
In Table~\ref{tab:obs-summary}, we list the effective characteristic baseline lengths, $\tilde{B}_\mathrm{c}$, and corresponding effective tied-array beam widths.
For our observations in the compact configuration, $\tilde{B}_\mathrm{c}\sim\qty{240}{\m}$, whereas in the extended configuration, $\tilde{B}_\mathrm{c}\sim\qty{3200}{\m}$.

The flexibility of the VCS data and the large field of view of the primary beam ($\sim$610\,$\mathrm{deg}^2$ near zenith at \qty{154}{\MHz}) together enable simultaneous observations of multiple pulsars by beamforming the tile voltages towards each position.
Taking advantage of this unique capability of the MWA, we observed the double pulsar as well as five bright non-recycled pulsars within the main lobe of the primary beam \citep[e.g.][]{Bhat2026}.
The characteristics of the observed pulsars are summarized in Table~\ref{tab:pulsars}.
Besides being useful for comparing scattering measurements, these secondary targets can be used to assess calibration quality and beam coherence between observing epochs.

\begin{deluxetable}{lcccc}
\tablewidth{0pt}
\tablecaption{Characteristics of the observed pulsars. \label{tab:pulsars}}
\tablehead{
    Name &
    \colhead{Period} &
    \colhead{DM} &
    \colhead{Distance} &
    \colhead{Ref.} \\[-5pt]
    (J2000)&
    \colhead{[\unit{\ms}]} &
    \colhead{[\unit{\per\cm\cubed\pc}]} &
    \colhead{[\unit{\kpc}]} &
}
\startdata
J0630$-$2834  & 1244.4 &  34.39 & $0.32^{+0.05}_{-0.04}$ & [1,2] \\
J0737$-$3039A &   22.7 &  48.91 & $0.74\pm0.06$ & [3] \\
J0742$-$2822  &  166.8 &  73.80 & $2.0^{+1.0}_{-0.8}$ & [1] \\
J0820$-$4114  &  545.4 & 113.34 & 0.32 & [5] \\
J0835$-$4510  &   89.3 &  67.73 & $0.28\pm0.02$ & [1,4] \\
J0837$-$4135  &  751.6 & 147.34 & $1.5^{+1.2}_{-0.9}$ & [1] \\
\enddata
\tablecomments{References for distances: [1] \citet{Verbiest2012}; [2] \citet{Deller2009}; [3] \citet{Kramer2021PhRvX}; [4] \citet{Dodson2003}; [5] DM-derived distance estimate from NE2025 \citep{NE2025}.}
\end{deluxetable}

\subsection{Tile amplitude and phase calibration}

We collected dedicated observations of standard calibrator sources using the MWAX correlator immediately before and after each VCS observation.
Each calibration observation was \qty{120}{\s} in duration and used the same channel configuration as the VCS observations.
The calibrator observations were pre-processed using \textsc{birli}, which performs flagging of known bad tiles and downsampling to a frequency/time resolution of \qty{40}{\kHz}/\qty{2}{\s}.
The pre-processed visibilities were then passed to \textsc{mwa\_hyperdrive} \citep{hyperdrive}, along with sky and beam models, to solve for the tile gains as a function of frequency.
Tiles with excessive variance in the gain amplitudes or phases, or significantly non-linear phase ramps, were manually flagged.
We also flagged \qty{80}{\kHz} at the top and bottom edges and \qty{40}{\kHz} at the center of each coarse channel, leaving a usable bandwidth of \qty{6.48}{\MHz} per subband.
For all observations, we found that the best calibration solutions came from observations of Pictor A.
For Epoch C, we could not determine a calibration solution in the \qty[number-unit-product=-]{179.84}{\MHz} subband using our dedicated calibration observation.
Instead, for this subband, we obtained a calibration solution using an observation of Centaurus A from $\sim$\qty{3}{\hour} after the end of the VCS observation.

\subsection{Data processing}

The VCS data were transferred to the Pawsey Supercomputing Centre in Perth, where the offline data processing was performed on the Setonix cluster.
We used \textsc{vcsbeam} \citep{Ord2019,Bhat2026} to apply the calibration solutions to the tile voltages, channelize the data to a frequency resolution of \qty{10}{\kHz}, and form tied-array beams towards each target pulsar.
The beamformed data were converted to Stokes parameters and output in the \textsc{psrfits} file format \citep{Hotan2004}.
The data were then folded with \textsc{dspsr} \citep{vanStraten2011} using the publicly-available folding ephemerides from the MeerTime project\footnote{\url{https://pulsars.org.au/}} \citep{Kramer2021MNRAS,Keith2024}.
For PSR J0737$-$3039A, we folded the data into 256 phase bins with \qty[number-unit-product=-]{1}{\s} subintegrations.
For the other pulsars, we used 512 phase bins and \qty[number-unit-product=-]{10}{\s} subintegrations.
The baseline was removed from each frequency channel in the folded data using \textsc{psrchive} \citep{vanStraten2012}.
We then used \textsc{clfd} to flag radio frequency interference (RFI) in the folded data using both profile masking\footnote{We found that the profile outlier identification performed the best using the \texttt{std}, \texttt{ptp}, and \texttt{lfamp} features.} and time-phase spike subtraction \citep{Morello2019}.
In general, the observations were exceptionally clean ($<\qty{1}{\percent}$ of data flagged).
A small amount of impulsive RFI was identified in the \qty[number-unit-product=-]{180}{\MHz} subband of Epoch B, but this was easily mitigated by \textsc{clfd} and affected $<\qty{3}{\percent}$ of the data for that subband.
After RFI flagging, the data were dedispersed and averaged to \qty[number-unit-product=-]{10}{\s} subintegrations using the \textsc{pam} program from \textsc{psrchive}.
By performing the dedispersion for J0737$-$3039A on \qty[number-unit-product=-]{1}{\s} subintegrations, we minimize the dispersive Doppler smearing effects described by \citet{Hu2025}.

%% file: 3-Analysis-Results.tex
\section{Analysis and Results}\label{sec:dp:analysis}

\begin{figure}[t]
    \centering
    \includegraphics[width=\linewidth]{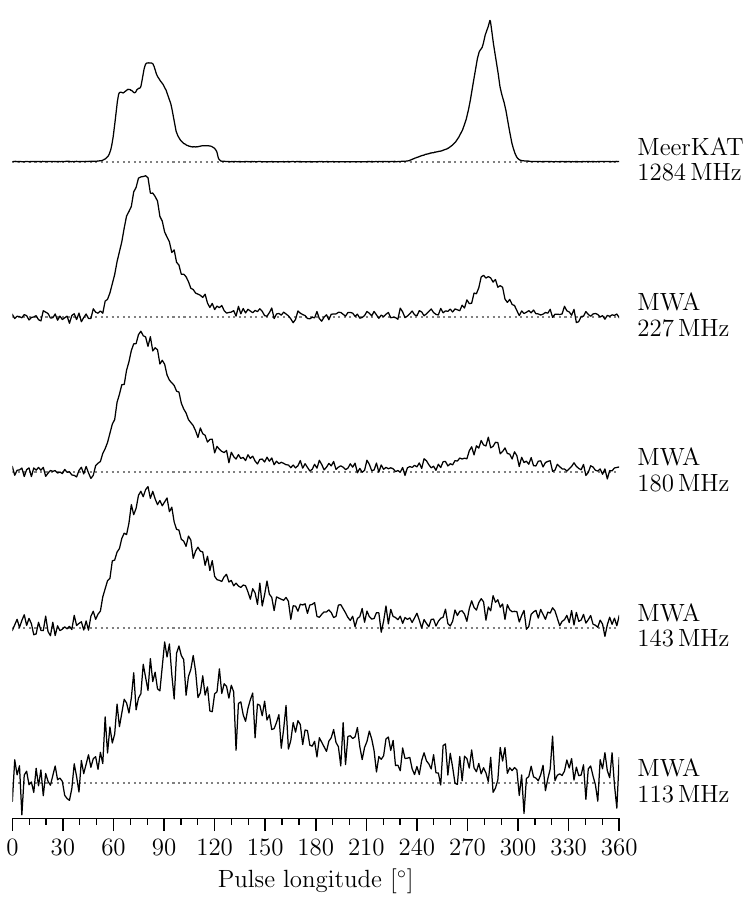}
    \caption{Frequency evolution of the integrated pulse profile of PSR J0737$-$3039A. The MWA profiles are averaged over Epochs A and B. For comparison, a MeerKAT $L$-band profile is shown \citep{Spiewak2022}. The MeerKAT profile was aligned with the MWA profiles based on the frequency evolution reported in \citet{Kramer2008}.}
    \label{fig:J0737-profiles}
\end{figure}

\subsection{Pulsar detections}

The six target pulsars in Table~\ref{tab:pulsars} were detected in all of the observing epochs and the pulse profiles are provided in Appendix~\ref{app:profiles}.
For these pulsars, the detections presented here are the highest quality ever published at low frequencies \citep[cf.][]{Geyer2017,Kirsten2019,Xue2019,XueThesis,Kumar2026}.
Additionally, the detections extend to the lowest frequencies ever published for J0737$-$3039A, J0820$-$4114, J0835$-$4510 (the Vela pulsar), and J0837$-$4135.
We note that the detections made in the extended array configurations (Epochs C and D) were less sensitive than the ones in the compact array configuration (Epochs A and B), which we attribute to a lower beamforming coherence.
The sensitivity in Epoch C is the lowest of the four observations due to a significant number of flagged tiles (see Table~\ref{tab:obs-summary}).
Nevertheless, all six pulsars were detected in Epoch C.

For the detections of J0737$-$3039A in Epochs A and B, the total subband-integrated profile S/N (reported by \textsc{psrchive}) ranges from $\sim$30 at \qty{113}{\MHz} to $\sim$90 at \qty{180}{\MHz}.
In \qty[number-unit-product=-]{10}{\s} subintegrations, the signal is detectable with a mean subband-integrated S/N of $\sim$6 in the top three subbands.
These observations are only sensitive to single-pulse emission at the $\gtrsim$\qty{20}{\Jy} level for an optimal single-pulse search with a limiting S/N of 5.

\subsection{DM measurements}\label{sec:res:dm}

To measure the DM, we used the template matching tools in \textsc{psrchive} \citep{vanStraten2012} to obtain pulse times of arrival (TOAs) at multiple frequencies, then fit a timing model with \textsc{tempo2} \citep{Tempo2}.
We started by co-adding the Epoch A and B detections and frequency-averaging to form one profile per \qty[number-unit-product=-]{3.84}{\MHz} subband with the highest possible S/N (see Figure~\ref{fig:J0737-profiles}).
Template profiles were then created by smoothing the profile from each subband using the wavelet fitting method in \textsc{psrsmooth}.
Pulse TOAs were measured for each subband with \qty[number-unit-product=-]{15}{\minute} spacing using the Fourier phase gradient method \citep{Taylor1992} with Markov chain Monte Carlo (MCMC) in \textsc{pat}, producing 80 TOAs per observation.
For Epoch C, we manually flagged a small number of TOAs that had very large uncertainties, indicating that the profiles were not of sufficiently high quality to obtain reliable TOAs.

The uncertainties in the measured TOAs increase substantially in the lower subbands due to pulse broadening and the lower S/N.
For Epochs A and B, the median TOA uncertainty is $\sim$\qty{50}{\us} at \qty{180}{\MHz} and \qty{227}{\MHz}, $\sim$\qtyrange{90}{100}{\us} at \qty{143}{\MHz}, and $\sim$\qtyrange{330}{350}{\us} at \qty{113}{\MHz}.
Due to the large uncertainties, we excluded TOAs in the \qtyrange[number-unit-product=-]{109}{117}{\MHz} subband from our DM analysis.
For comparison, the DMs were measured twice using the TOAs from the three and two highest-frequency subbands, and the results are summarized in Table~\ref{tab:dm}.
The DMs measured over the top three subbands are systematically higher by $\sim$\qty{e-3}{\per\cm\cubed\pc} than when measured over the top two subbands.
This difference is most likely due to pulse broadening.
Therefore, although the frequency span is smaller, we expect that the DMs measured over the top two subbands are likely more reliable.

\begin{deluxetable}{lcc}[t]
\tablewidth{0pt}
\tablecaption{DM measurements for PSR J0737$-$3039A from \textsc{tempo2} fits to different subsets of TOAs.\label{tab:dm}}
\tablehead{Epoch(s) & \multicolumn{2}{c}{DM [\unit{\per\cm\cubed\pc}]} \\[2pt]
\cline{2-3}
& \colhead{\qtyrange{139}{231}{\MHz}} & \colhead{\qtyrange{176}{231}{\MHz}}}
\startdata
A      & 48.9118(2) & 48.9109(4) \\
B      & 48.9117(2) & 48.9109(4) \\
A + B  & 48.9117(2) & 48.9109(3) \\
\hline
C      & 48.9152(7) & 48.913(1)  \\
D      & 48.9135(5) & 48.9126(8) \\
C + D  & 48.9138(4) & 48.9127(6) \\
\enddata
\end{deluxetable}

\subsection{Pulse Broadening Measurements}\label{sec:pb-method}

Our pulse broadening analysis uses a forward-modeling approach, where we assume that the intrinsic pulse profile, $p(t)$, can be described by the sum of $N$ Gaussian components,
\begin{equation}
    p(t) = \sum_{i=1}^{N}\frac{F_i}{\sigma_i\sqrt{2\pi}}\exp\left[-\frac{1}{2}\left(\frac{t-t_{0,i}}{\sigma_i}\right)^2\right],
\end{equation}
where $t_{0,i}$, $\sigma_i$, and $F_i$ define the location, width, and fluence of the $i$-th component \citep{McKinnon2014,Jankowski2023}.
The observed pulse profile is the convolution of the intrinsic pulse profile with pulse broadening functions (PBFs) describing the effects of multipath scattering in the IISM, $g(t)$, and dispersive smearing, $d(t)$:
\begin{equation}
    \tilde{p}(t) = b + p(t) * g(t) * d(t),
\end{equation}
where $*$ denotes linear convolution \citep{Loehmer2001}.
The parameter $b$ is a baseline offset on the order of the noise level to account for any error in the baseline subtraction \citep{Geyer2016}.
We adopt the thin-screen scattering model for which the PBF is a single-sided exponential,
\begin{equation}
    g(t) = \frac{1}{\tau_\mathrm{s}}\exp\left(-\frac{t}{\tau_\mathrm{s}}\right)H(t),
\end{equation}
where $H(t)$ is the Heaviside step function and $\tau_\mathrm{s}$ is the pulse broadening timescale \citep{Williamson1972}.
Since the channel bandwidth (\qty{10}{\kHz}) is small compared with the center frequency, the dispersion sweep within a single channel is approximately linear and $d(t)$ is a rectangular function with a width equal to the dispersive delay over the channel, $\tau_\mathrm{DM}$\footnote{For a small channel bandwidth, the dispersive smearing scales as $\tau_\mathrm{DM}\propto\nu^{-3}$ \citep[e.g.\ see Section~A2.4 of][]{LorimerKramer2012}.}.
In this case, we can assume that convolution with $d(t)$ causes symmetrical broadening of $p(t)$ that is effectively indistinguishable from an increase in $\sigma$ of each component.
Therefore, for simplicity, we ignore $d(t)$ and interpret $\sigma$ as a combination of the intrinsic pulse component width and the dispersive smearing.

\begin{deluxetable}{lcccc}[t]
\tablewidth{0pt}
\tablecaption{Measurements of the pulse broadening timescale $\tau_\mathrm{s}$ from forward modeling of pulse profiles.\label{tab:tau}}
\tablehead{Epoch & \multicolumn{4}{c}{$\tau_\mathrm{s}$ [\unit{\ms}]} \\[2pt]
\cline{2-5}
& \colhead{\qty{113}{\MHz}} & \colhead{\qty{143}{\MHz}} & \colhead{\qty{180}{\MHz}} & \colhead{\qty{227}{\MHz}}}
\startdata
\multicolumn{5}{c}{PSR J0737$-$3039A} \\
\hline
A    & 5.2(6)   & 2.7(1) & 1.40(5) & 0.84(4) \\
B    & 5.9(6)   & 2.6(1) & 1.32(6) & 0.91(3) \\
C    & \nodata  & 3.0(8) & 2.4(5)  & 1.2(2)  \\
D    & 7(2)     & 3.0(4) & 1.9(1)  & 0.99(9) \\
Avg. & 5.6(5)   & 2.7(1) & 1.4(2)  & 0.90(5) \\
\hline
\multicolumn{5}{c}{PSR J0742$-$2822} \\
\hline
A    & 33.5(9)   & 20.6(3) & 6.3(1)  & 3.30(3) \\
B    & 32.2(7)   & 20.1(2) & 6.52(8) & 3.27(2) \\
C    & 41(4)     & 22.5(8) & 6.9(4)  & 3.8(1)  \\
D    & 39(1)     & 13.9(5) & 6.59(8) & 3.46(4) \\
Avg. & 34(2)     & 20(2)   & 6.5(1)  & 3.32(9) \\
\hline
\multicolumn{5}{c}{PSR J0837$-$4135} \\
\hline
A    & 62(4)   & 23.1(7)   & 9.2(2)  & 3.2(2) \\
B    & 64(5)   & 27.6(8)   & 10.7(3) & 3.5(2) \\
C    & \nodata & 25(5)     & 7(1)    & 4.0(7) \\
D    & \nodata & 26(2)     & 9.9(5)  & 5.0(5) \\
Avg. & 62.9(7) & 25(2)     & 9.7(8)  & 3.5(4) \\
\enddata
\tablecomments{The individual measurements are the medians and 1-$\sigma$ credible intervals of the posterior distributions. The average measurements are the weighted means and weighted standard deviations.}
\end{deluxetable}

To perform the model fitting, we used the \textsc{scatfit} Python package \citep{scatfit}.
\textsc{scatfit} was originally developed for analysis of Fast Radio Bursts \citep[e.g.][]{Jankowski2023}, but it is also capable of analyzing folded pulsar data so long as the scattering tail does not wrap around in phase and an off-pulse region can be unambiguously defined.
In cases where the scattering tail extends over more than one period, the increase in baseline must be accounted for using the ``DC + train" method described by \citet{Geyer2016} or similar.
The profile model fitting was performed independently for each subband.
For each fit, \textsc{scatfit} determines an initial set of parameter values using least-squares optimization.
The fit is then refined using MCMC sampling, producing a distribution of posterior samples, and the convergence is assessed using standard methods.
If the fit is successful, the pulse width at half maximum for the intrinsic Gaussian components and the total scattered profile ($W_\mathrm{50i}$ and $W_\mathrm{50p}$, respectively) are calculated as described in \citet{Jankowski2023}.

\begin{figure*}[p]
    \centering
    \includegraphics[width=\linewidth]{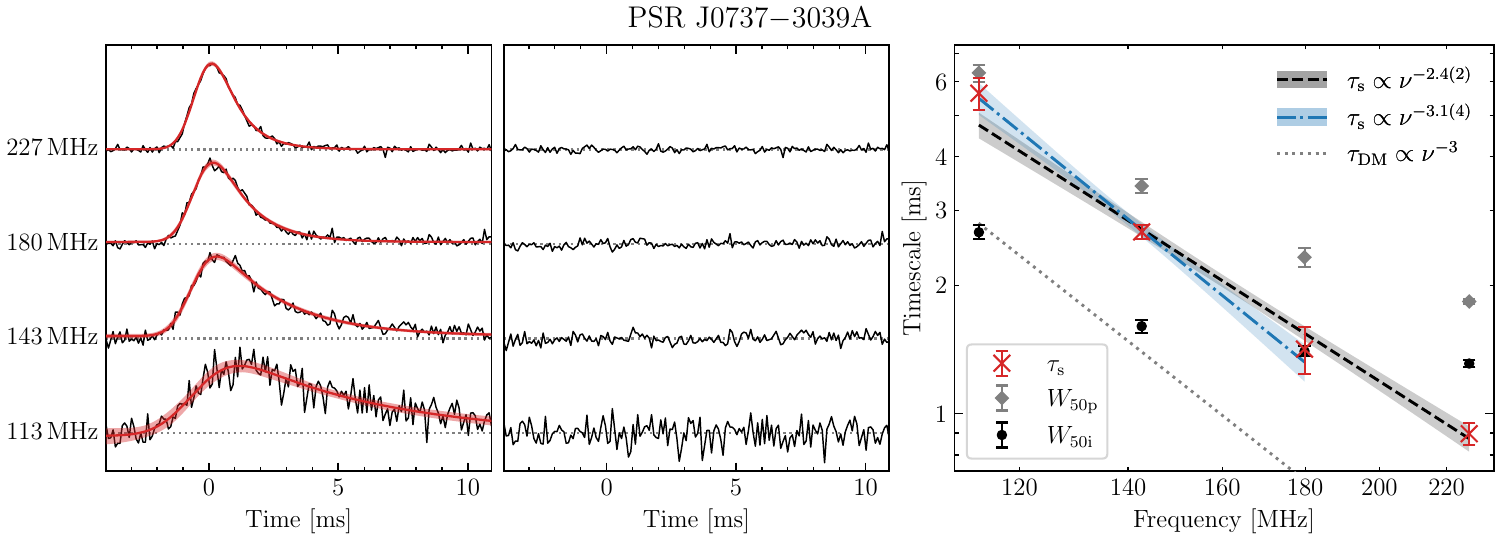}
    \includegraphics[width=\linewidth]{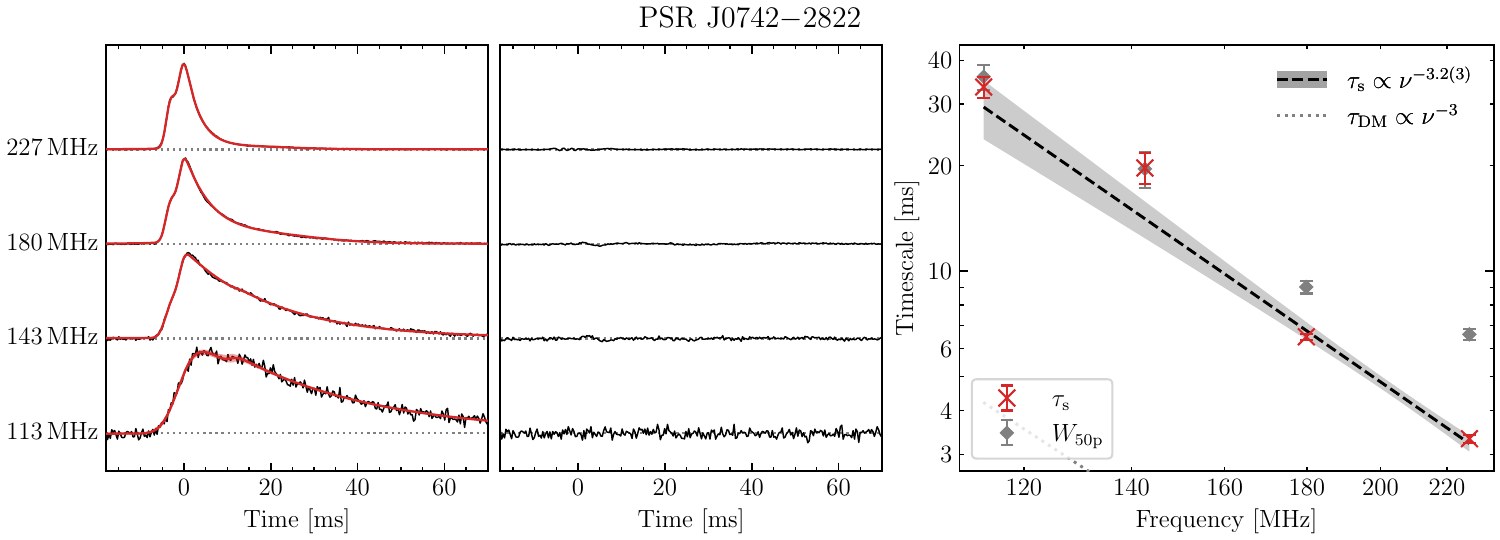}
    \includegraphics[width=\linewidth]{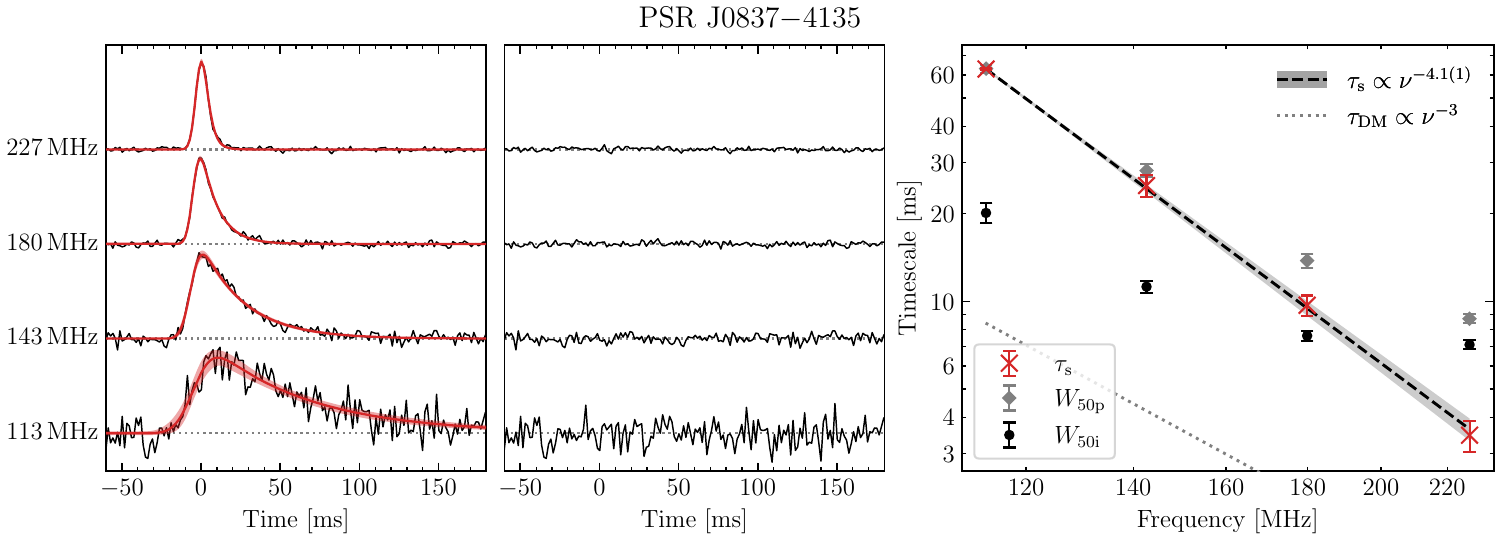}
    \caption{Pulse broadening model fits for PSR J0737$-$3039A (top), PSR J0742$-$2822 (middle), and PSR J0837$-$4135 (bottom). \textit{Left panels}: Pulse profiles for each subband from Epoch B (black lines). The red lines show the median model fits and the shaded red bands are the 3-$\sigma$ credible intervals. \textit{Center panels}: The fit residuals, i.e.\ the difference between the data and the median model. \textit{Right panels}: Spectrum of pulse broadening timescales ($\tau_\mathrm{s}$; red crosses), scattered pulse widths ($W_\mathrm{50p}$; grey diamonds), and intrinsic pulse widths ($W_\mathrm{50i}$; black circles) averaged over all epochs. The $W_\mathrm{50i}$ estimates are not shown for J0742$-$2822 as the profile components vary between epochs due to mode changing. The dispersive smearing ($\tau_\mathrm{DM}$) is indicated by the dotted line. The dashed black line and blue dash-dotted line show the best-fit power-law spectrum to $\tau_\mathrm{s}$ over all subbands and just the lower three subbands, respectively. The shaded bands are the 1-$\sigma$ uncertainties in the power-law fits.}
    \label{fig:scatfit}
\end{figure*}

Table~\ref{tab:tau} summarizes the posterior estimates of $\tau_\mathrm{s}$ for J0737$-$3039A and the two non-recycled pulsars for which reliable scattering fits could be obtained, J0742$-$2822 and J0837$-$4135.
In the cases where satisfactory convergence could not be obtained (which was an issue only for Epoch C due to low S/N), we have excluded those measurements.
The profile fits for Epoch B are shown in the left panels of Figure~\ref{fig:scatfit}.

Having multiple measurements separated by months provides the opportunity of being able to average over long-term variations in the pulse broadening timescale that occur due to refractive interstellar scintillation \citep[RISS; e.g.][]{Rickett1990}.
The RISS decorrelation timescale is related to the DISS decorrelation timescale and bandwidth as $\Delta t_\mathrm{r}=\nu\Delta t_\mathrm{d}/\Delta\nu_\mathrm{d}$ \citep{LorimerKramer2012}.
From the \textit{UHF}-band scintillation measurements of J0737$-$3039A reported by \citet{Askew2026}, we find $\Delta t_\mathrm{r}\approx\qty{8}{\day}$.
Assuming a Kolmogorov spectrum, the frequency scaling is $\Delta t_\mathrm{r}\propto\nu^{-2.2}$.
Therefore, $\Delta t_\mathrm{r}$ is approximately 18, 11, 7, and 4 months from lowest to highest frequency for the four MWA subbands (Table~\ref{tab:obs-setup}).
Although the total span of the four MWA observations is 14 months, the sampling is very sparse.
To fully average over RISS variations would require sampling at a roughly monthly cadence over several RISS cycles.
Nonetheless, we effectively sample two RISS cycles for all except the lowest subband.
The averaged measurements should therefore be marginally more reliable than the single-epoch measurements.
In Table~\ref{tab:tau}, we report the weighted mean and the weighted standard deviation (stated as the uncertainty) of $\tau_\mathrm{s}$ for each pulsar and subband.
This weighted standard deviation quantifies the intrinsic variance of the measurements that is expected due to RISS.

To measure the spectral index, $\alpha$, we fit a power-law,
\begin{equation}
    \tau_\mathrm{s}(\nu) = \tau_0 \left(\frac{\nu}{\nu_0}\right)^\alpha,
\end{equation}
where $\tau_0$ is the pulse broadening timescale at the reference frequency $\nu_0$ (chosen to be \qty{150}{\MHz}).
For a power-law spectrum, the effective frequency of each $\tau_\mathrm{s}$ measurement is the geometric mean of the subband frequency range \citep[see][for details]{Geyer2016}.
It is also important to consider that the model does not account for subband-dependent systematic errors that may occur due to unmodeled profile complexity or insufficient data quality.
We therefore opt for robust least-squares fitting, which places less weight on measurements that deviate far from the mean model.
The fitting was performed with the \textsc{iminuit} Python library \citep{minuit,iminuit} using the \texttt{soft\_l1} loss function\footnote{For details, see \url{https://docs.scipy.org/doc/scipy/reference/generated/scipy.optimize.least_squares.html}.} and the \textsc{migrad} minimization algorithm.
The fits were performed on the epoch-averaged pulse broadening measurements and the results are summarized in Table~\ref{tab:alpha}.
For each fit, the reduced $\chi^2$ is provided as a quality-of-fit statistic.
The quoted uncertainties are the standard errors obtained from the square root of the diagonal elements of the covariance matrix.
The spectra are shown in the right panels of Figure~\ref{fig:scatfit}.

\begin{deluxetable}{lcccc}
\tablewidth{0pt}
\tablecaption{Results of power-law fits to $\tau_\mathrm{s}(\nu)$ using measurements from this work.\label{tab:alpha}}
\tablehead{
    PSR &
    \colhead{$\nu$ range} &
    \colhead{$\alpha$} &
    \colhead{$\tau_\mathrm{s}$ at \qty{150}{\MHz}} &
    \colhead{$\chi_\mathrm{r}^2$} \\
    &
    \colhead{[\unit{\MHz}]} &
    &
    \colhead{[\unit{\ms}]} &
}
\startdata
J0737$-$3039A & 109--184 & $-2.4(2)$ & $2.4(1)$  & 1.5 \\
              & 109--231 & $-3.1(4)$ & $2.3(1)$   & 0.4 \\
J0742$-$2822  & 109--231 & $-3.2(3)$ & $12(1)$    & 4.9 \\
J0837$-$4135  & 109--231 & $-4.1(1)$ & $19.9(7)$  & 0.2 \\
\enddata
\end{deluxetable}

\subsubsection{PSR J0737\safeminus{}3039A}\label{sec:scat-results-0737}

The leading (main) pulse in J0737$-$3039A's profile has only one clearly distinguishable component below \qty{300}{\MHz}.
The trailing (inter)-pulse has a shallow precursor component and a larger main component, making it more complex to model.
Since the interpulse also has a lower S/N and is only detectable in the top three subbands, we have excluded it from the model fit.
We model the leading pulse with a single component because the addition of a second component (although empirically motivated from higher-frequency profiles) may lead to overfitting given the S/N.
Over the four subbands, we measure $\alpha=-2.4(2)$.
However, the measurements deviate from a power law (see Figure~\ref{fig:scatfit}), indicating that there are systematic errors present.
This could be due to the trailing component of the main pulse, which can be seen in the MeerKAT profile in Figure~\ref{fig:J0737-profiles}.
Below $\sim$\qty{400}{\MHz}, the components in the main pulse merge, and the trailing component mimics the shape of a scattering tail \citep[cf.\ Figure~3 of][]{Kramer2008}.
It is possible that the unmodeled component may be contributing to an overestimation of $\tau_\mathrm{s}$ at \qty{227}{\MHz}.
This is also the only subband where $W_\mathrm{50i}$ significantly exceeds $\tau_\mathrm{s}$, making the fit more sensitive to the accuracy of the intrinsic profile model.
Excluding the measurement at \qty{227}{\MHz}, the frequency scaling of $\tau_\mathrm{s}$ is consistent with a power law with $\alpha=-3.1(4)$ (as reported in Table~\ref{tab:alpha}).
This second fit also has a smaller $\chi_\mathrm{r}^2$ of 0.4 compared with 1.5 for the fit to all subbands, indicating that the data are more consistent with the model.

\subsubsection{PSR J0742\safeminus{}2822}

Visual inspection of J0742$-$2822's profile at higher frequencies indicates that it contains roughly six components that vary in amplitude between two emission modes \citep{Keith2013}.
Interestingly, although our observations are \qty{2.5}{\hour} each ($\sim$\num{3.3e6} pulses), the integrated profile is visibly different in each observation.
This indicates that the emission modes persist on timescales comparable to our observation duration ($\sim$hours).\footnote{Mode changing for this pulsar has not previously been studied in long continuous observations, making our data particularly unique.
A more detailed analysis of these observations, including mode separation and characterization, is deferred to a future publication.}
As such, individual observations appear to be dominated by one of the two modes.
The number of detectable components also varies between subbands due to the intrinsic profile evolution.
In a previous pulse broadening analysis at \qty{327}{\MHz} by \citet{Krishnakumar2015}, only the trailing pulse component was modeled.
However, to reduce the bias due to having an incomplete model of the intrinsic profile, we chose to model all of the detectable components.
For each profile, the number of components was chosen based on the convergence of the fit (e.g.\ by inspecting the corner plot and Markov chains), the flatness of the residuals, and various quality-of-fit statistics including the reduced $\chi^2$ and the Akaike information criterion (AIC).
In general, we found that there are three components: a leading component that varies in amplitude and is detectable only in the top three subbands, a central component that is relatively stable between observations, and a trailing component that appears only in some observations (see Appendix~\ref{app:profiles} for the profiles from each epoch).

According to the best-fit model, the \qty[number-unit-product=-]{143}{\MHz} measurement is an outlier, being outside of the 1-$\sigma$ error band.
This discrepancy is seen in all epochs, even when different numbers of components are visible.
Although the source of the error is unknown, it highlights the main limitation of the forward modeling approach: if the intrinsic profile cannot be accurately modeled, then the unmodeled complexity will be absorbed into other fit parameters, resulting in systematic errors.
Having independent measurements from multiple subbands can reveal these errors.
Based on the larger uncertainty and distance from the mean power-law model, the \qty[number-unit-product=-]{143}{\MHz} measurement of $\tau_\mathrm{s}$ appears to be somewhat overestimated.
The accuracy of the forward modeling may be able to be improved by separating the emission modes.
The mode-separated pulse profiles could then be co-added over multiple observing epochs to increase the S/N and develop a better characterization of the intrinsic pulse components.

\subsubsection{PSR J0835\safeminus{}4510 (the Vela pulsar)}

The profile of this pulsar is difficult to model due to the scattering tail extending over more than one pulse period.
As mentioned in Section~\ref{sec:pb-method}, to correctly model this data would require the application of a more complicated method that accounts for the raised baseline and distorted profile \citep[e.g.][]{Geyer2016}.
A careful analysis of pulse broadening for these detections will be deferred to a future publication.

\subsubsection{PSR J0837\safeminus{}4135}

The profile of J0837$-$4135 is well modeled by a single Gaussian and it shows substantial evolution in pulse broadening over the observed subbands.
These features make it an ideal pulsar for forward modeling the pulse profile.
The $\tau_\mathrm{s}$ spectrum is well described by a power law with $\alpha=-4.1(1)$ (see Figure~\ref{fig:scatfit}).

\subsubsection{Pulsars without scattering measurements}
\paragraph{PSR J0630$-$2834} This pulsar does not show any visible pulse broadening, even at the lowest frequencies of our observations.
This is unsurprising because it is located outside of the Gum Nebula (see Figure~\ref{fig:hmap}).
Given that this is a bright pulsar with a low DM of \qty{34.4}{\per\cm\cubed\pc} and relatively long period of \qty{1.2}{\s}, scintillation studies may be possible in future.

\paragraph{PSR J0820$-$4114} Although the trailing side of the profile exhibits a decay that resembles a scattering tail (see Appendix~\ref{app:profiles}), the pulse shape does not evolve significantly over the observed frequencies.
This suggests that the tail is an intrinsic feature of the profile.
If there is scattering, it is difficult to constrain due to the low S/N and the frequency evolution of the relatively complex intrinsic profile \citep[e.g.][]{McSweeney2023}.

%% file: 4-Discussion.tex
\section{Discussion}\label{sec:dp:discussion}

\subsection{Joint analysis of pulse broadening and scintillation measurements}\label{sec:joint}

\begin{deluxetable}{lccccccc}
\tablewidth{0pt}
\tablecaption{Results of power-law fits to $\tau_\mathrm{s}(\nu)$ when scintillation data are included using different assumptions of the proportionality constant $C_1$ in Equation~\eqref{eq:scat-scin-relation}. In the last fit for each pulsar, $C_1$ is included as a free parameter in the model.\label{tab:alpha-lit}}
\tablehead{
    PSR &
    \colhead{Ref.} &
    \colhead{$\nu$ range} &
    \colhead{$C_1$} &
    \colhead{$\alpha$} &
    \colhead{$\tau_\mathrm{s}$ at \qty{150}{\MHz}} &
    \colhead{$\tau_\mathrm{s}$ at \qty{1}{\GHz}} &
    \colhead{$\chi_\mathrm{r}^2$}\\
    &
    &
    \colhead{[\unit{\GHz}]} &
    &
    &
    \colhead{[\unit{\ms}]} &
    \colhead{[\unit{\us}]} &
}
\startdata
J0737$-$3039A & [1] & 0.11--2.62 & 0.654 & $-4.3(1)$   & $1.9(4)$  & $0.5(1)$  & 0.8 \\
              &     &            & 1.000 & $-4.1(1)$   & $2.0(4)$  & $0.8(2)$  & 0.6 \\
              &     &            & 4(4)  & $-3.6(4)$   & $2.3(4)$  & $2.8(8)$  & 0.3 \\
\hline
J0742$-$2822  & [2] & 0.11--9.00 & 0.654 & $-4.0(1)$   & $13(3)$   & $7(1)$    & 0.4 \\
              &     &            & 1.000 & $-3.8(1)$   & $13(2)$   & $9(2)$    & 0.2 \\
              &     &            & 2(2)  & $-3.6(4)$   & $14(2)$   & $14(8)$   & 0.2 \\
\enddata
\tablecomments{Scintillation data: [1] \citet{Askew2026}; [2] \citet{Wang2025}.}
\end{deluxetable}

\begin{figure}
    \centering
    \includegraphics[width=\linewidth]{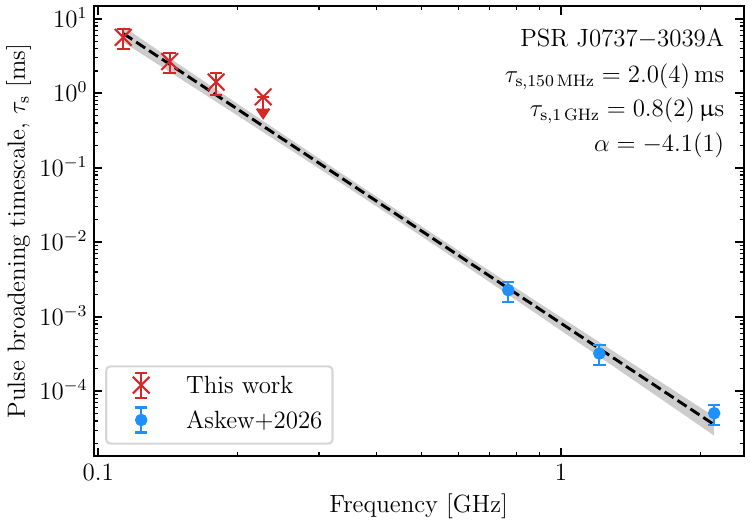}
    \includegraphics[width=\linewidth]{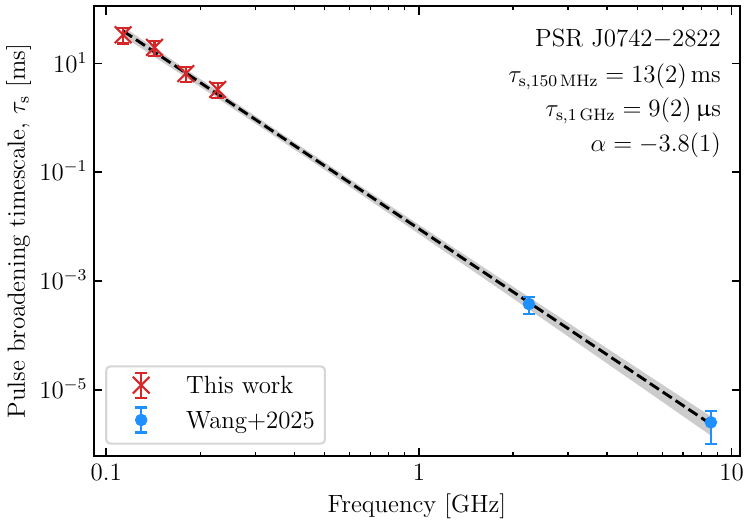}
    \caption{Combined power-law fits to the MWA $\tau_\mathrm{s}$ measurements from this work (red crosses) and $\tau_\mathrm{s}$ measurements derived from published scintillation bandwidths assuming $C_1=1$ (blue circles). \textit{Top}: Spectrum of PSR J0737$-$3039A using data from \citet{Askew2026}. \textit{Bottom}: Spectrum of PSR J0742$-$2822 using data from \citet{Wang2025}. A \qty{30}{\percent} uncertainty has been added to all measurements to account for systematic errors. The dashed lines show the best-fit models and the gray shaded bands show the 1-$\sigma$ fit uncertainties.}
    \label{fig:tau-spectrum}
\end{figure}

The Fourier uncertainty relation defined in Equation~\eqref{eq:scat-scin-relation} enables us to convert measurements of the scintillation bandwidth, $\Delta\nu_\mathrm{d}$, into equivalent measurements of the pulse broadening timescale, $\tau_\mathrm{s}$.
Under the right circumstances, the proportionality constant $C_1$ can be determined using simultaneous measurements of $\tau_\mathrm{s}$ and $\Delta\nu_\mathrm{d}$ \citep[e.g.][]{Marthi2025}.
In the absence of a direct measurement, we assume values derived by \citet{Lambert1999} for a thin-screen geometry and a Kolmogorov wavenumber spectrum ($\alpha=-4.4$, $C_1=0.654$) or a square-law wavenumber spectrum ($\alpha=-4$, $C_1=1$).

Out of the three pulsars for which we have obtained new measurements of $\tau_\mathrm{s}$, two of them have $\Delta\nu_\mathrm{d}$ measurements available at higher frequencies.
For J0737$-$3039A, \citet{Askew2026} analyzed 16 observations with the MeerKAT telescope spanning 18 months.
Their observations were collected simultaneously in the \textit{UHF} band (\qtyrange{0.54}{1.09}{\GHz}), $L$ band (\qtyrange{0.86}{1.71}{\GHz}), and $S$ band (\qtyrange{1.75}{2.62}{\GHz}).
They report mean measurements of $\Delta\nu_\mathrm{d}$ for each frequency band.
For J0742$-$2822, \citet{Wang2025} analyzed 21 observations with the Tian-ma Radio Telescope (TMRT) spanning almost 2 years that were simultaneously collected in the \qtyrange{2.2}{2.3}{\GHz} and \qtyrange{8.2}{9.0}{\GHz} frequency bands.
They report measurements of $\Delta\nu_\mathrm{d}$ for each frequency band and epoch, from which we calculated the weighted mean and weighted standard deviation for each frequency band.
Both of these data sets have sufficient time spans and enough measurements to average over most of the variations due to RISS.

We performed a joint analysis using the epoch-averaged pulse broadening measurements from Table~\ref{tab:tau} in combination with measurements calculated from the scintillation data described above.
As an attempt to account for systematic errors (both known and unknown), we added a \qty{30}{\percent} relative uncertainty in quadrature with all of the measurements.
For the reasons discussed in Section~\ref{sec:scat-results-0737}, we assumed that the \qty[number-unit-product=-]{227}{\MHz} measurement is overestimated for J0737$-$3039A, and excluded it from the fit.
As usual, since $\tau_\mathrm{s}$ is assumed to scale as a power law, the effective frequency of each measurement is the geometric mean of the frequency range of the data.
We performed power-law model fits to the combined data sets using robust least-squares fitting as described in Section~\ref{sec:pb-method}.
The results for the two different values of $C_1$ are presented in Table~\ref{tab:alpha-lit}, and the fits for $C_1=1$ are illustrated in Figure~\ref{fig:tau-spectrum}.
We also attempted to include $C_1$ as a free parameter in the model, but found that the strong correlation between $C_1$ and $\alpha$ resulted in poor constraints (see Table~\ref{tab:alpha-lit}).
Additionally, the best-fit values of $C_1$ are much higher than predicted by theoretical models \citep[$0.5\lesssim C_1\lesssim 1.5$; cf.][]{Cordes1998,Lambert1999}.
This may be because $C_1$ absorbs the in-band systematic biases in $\alpha$ that we previously identified in Section~\ref{sec:pb-method}.
Therefore, our interpretation in the following sections is based on the model fits where $C_1$ is fixed to the theoretical values.

For J0737$-$3039A, we measure $\alpha=-4.3(1)$ for $C_1=0.654$, which is consistent with the theoretical value for a Kolmogorov spectrum ($-4.4$).
Likewise, we measure $\alpha=-4.1(1)$ for $C_1=1$, which is consistent with the theoretical value for a square-law spectrum ($-4$).
Due to the correlation between $\alpha$ and $C_1$, the results are consistent with both a Kolmogorov and square-law spectrum.
It is however notable that the $\chi_\mathrm{r}^2$ values show a marginal preference for the square-law model.
On the other hand, for J0742$-$2822, we measure $\alpha=-4.0(1)$ for $C_1=0.654$ and $\alpha=-3.8(1)$ for $C_1=1$.
These measurements are consistent with each other and deviate significantly from a Kolmogorov spectrum.
The $\chi_\mathrm{r}^2$ values also show a preference for the square-law model.
The reduced sensitivity of $\alpha$ to the value of $C_1$ for J0742$-$2822 is due to the larger fractional frequency span of the data set.

\subsection{Comparisons with published measurements}

\subsubsection{PSR J0737\safeminus{}3039A}\label{sec:dis:0737}

The first measurement of $\alpha$ for J0737$-$3039A was from \citet{Askew2026}, who found $\alpha=-3.61(5)$ over the MeerKAT frequency bands (\qtyrange{0.54}{2.62}{\GHz}).
This value is unexpectedly shallow compared with the theoretical expectations.
In Section~\ref{sec:scat-results-0737}, we measured $\alpha=-3.1(4)$ from our MWA data, which is even further from the expectations.
However, our joint analysis in Section~\ref{sec:joint}, which spans a total frequency range of \qtyrange{0.11}{2.62}{\GHz}, indicates that $-4.4\lesssim\alpha\lesssim-4$, with the specific value depending on the assumed $C_1$.
This is consistent with theoretical expectations, which suggests that there are frequency-dependent systematic errors present in the individual data sets.
For the MWA measurements, the main systematics are likely to be from the model of the intrinsic profile and RISS.

\subsubsection{PSR J0742\safeminus{}2822}\label{sec:dis:0742}

There have been several previous studies of the $\tau_\mathrm{s}$ spectrum of J0742$-$2822 that we can compare our measurements with.
\citet{Lewandowski2015b} compiled $\tau_\mathrm{s}$ measurements from the published literature, supplemented with their own measurements from pulse profiles available from public databases (together spanning \qtyrange{102}{1518}{\MHz}) and measured $\alpha=-2.5(3)$.
However, based on the pulse broadening observed in our data and visual inspection of higher-frequency pulse profiles, it seems very unlikely that any scattering measurements obtained from forward modeling of the profile above $\sim$\qty{300}{\MHz} could be considered reliable.
Additionally, the few measurements below \qty{300}{\MHz} that were available at the time were from single epochs and were not simultaneous.
The combination of these factors places low confidence in their spectral fit.

\citet{Geyer2017} measured $\alpha=-3.8(4)$ and $\tau_\mathrm{s}=\qty{20(2)}{\ms}$ at \qty{150}{\MHz} using a LOFAR observation spanning \qtyrange{119}{179}{\MHz}.
Their spectral index measurement is consistent with our measurements from the joint analysis in Section~\ref{sec:joint}.
We measure a lower $\tau_\mathrm{s}$, but this could easily be explained by long-term variations in scattering.
\citet{Kirsten2019} measured $\alpha=-2.5(6)$ from an MWA-VCS observation between \qtyrange{170}{230}{\MHz} using the \textsc{clean}-based deconvolution method \citep{Bhat2003}.
Their spectral index is consistent with our in-band estimate of $-3.2(3)$, but is inconsistent with our steeper joint fit estimates.
It is interesting that their analysis, which used a different method that makes no a priori assumptions about the intrinsic profile, also produced a shallow spectrum.
The interpretation of their result is limited by the low S/N of the profiles and the smaller frequency span of their data.
A similar analysis of our higher-quality data set using a \textsc{clean}-based approach would be insightful to determine a robust in-band estimate of the spectral index that depends on fewer assumptions.

\subsubsection{PSR J0837\safeminus{}4135}

For this pulsar, \citet{Lewandowski2015b} reported $\alpha=-2.36(9)$ using four $\tau_\mathrm{s}$ measurements from the literature: two at \qty{160}{\MHz} and two at higher frequencies (above $\qty{600}{\MHz}$).
However, for the same reasons as we discussed in Section~\ref{sec:dis:0742}, we do not expect the $\tau_\mathrm{s}$ measurements from forward modeling to be reliable above $\sim$\qty{300}{\MHz}.
For example, \citet{Mitra2001} measured $\tau_\mathrm{s}=1(1)\,\unit{\ms}$ from an observation at \qty{327}{\MHz}.
Therefore, ours is likely the first precise and reliable measurement of the $\tau_\mathrm{s}$ spectrum for this pulsar.

\subsection{Scattering strength}\label{sec:scatt-strength}

\begin{deluxetable}{lcccc}
\tablewidth{0pt}
\tablecaption{Results of the scattering analysis assuming Kolmogorov turbulence. The parameter estimates are the medians and \qty{68}{\percent} confidence intervals of the probability distributions.\label{tab:sm}}
\tablehead{
    PSR &
    \colhead{$\Delta\nu_\mathrm{d}$ at \qty{150}{\MHz}} &
    \colhead{$D$} &
    \colhead{$\left\langle C_{n_\mathrm{e}}^2 \right\rangle$} &
    \colhead{SM} \\
    &
    \colhead{[\unit{\Hz}]} &
    \colhead{[\unit{\kpc}]} &
    \colhead{[$10^{-3}\,\mathrm{m}^{-20/3}$]} &
    \colhead{[$10^{-3}\,\mathrm{m}^{-20/3}\,\mathrm{kpc}$]}
}
\startdata
J0737$-$3039A & $55^{+12}_{-8}$ & $0.74\pm0.06$ & $11.8^{+2.7}_{-2.3}$ & $8.7^{+1.5}_{-1.4}$ \\
J0742$-$2822  & $7.8^{+1.8}_{-1.2}$ & $2.0^{+1.0}_{-0.7}$ & $10^{+11}_{-5}$ & $19^{+9}_{-6}$ \\
J0837$-$4135  & $5.23^{+0.19}_{-0.18}$ & $1.5^{+1.2}_{-0.7}$ & $23^{+44}_{-15}$ & $34^{+22}_{-13}$ \\
\enddata
\tablecomments{The distance uncertainties for J0742$-$2822 and J0837$-$4135 are different to the published values as we have modeled them with log-normal distributions.}
\end{deluxetable}

\begin{figure}
    \centering
    \includegraphics[width=\linewidth]{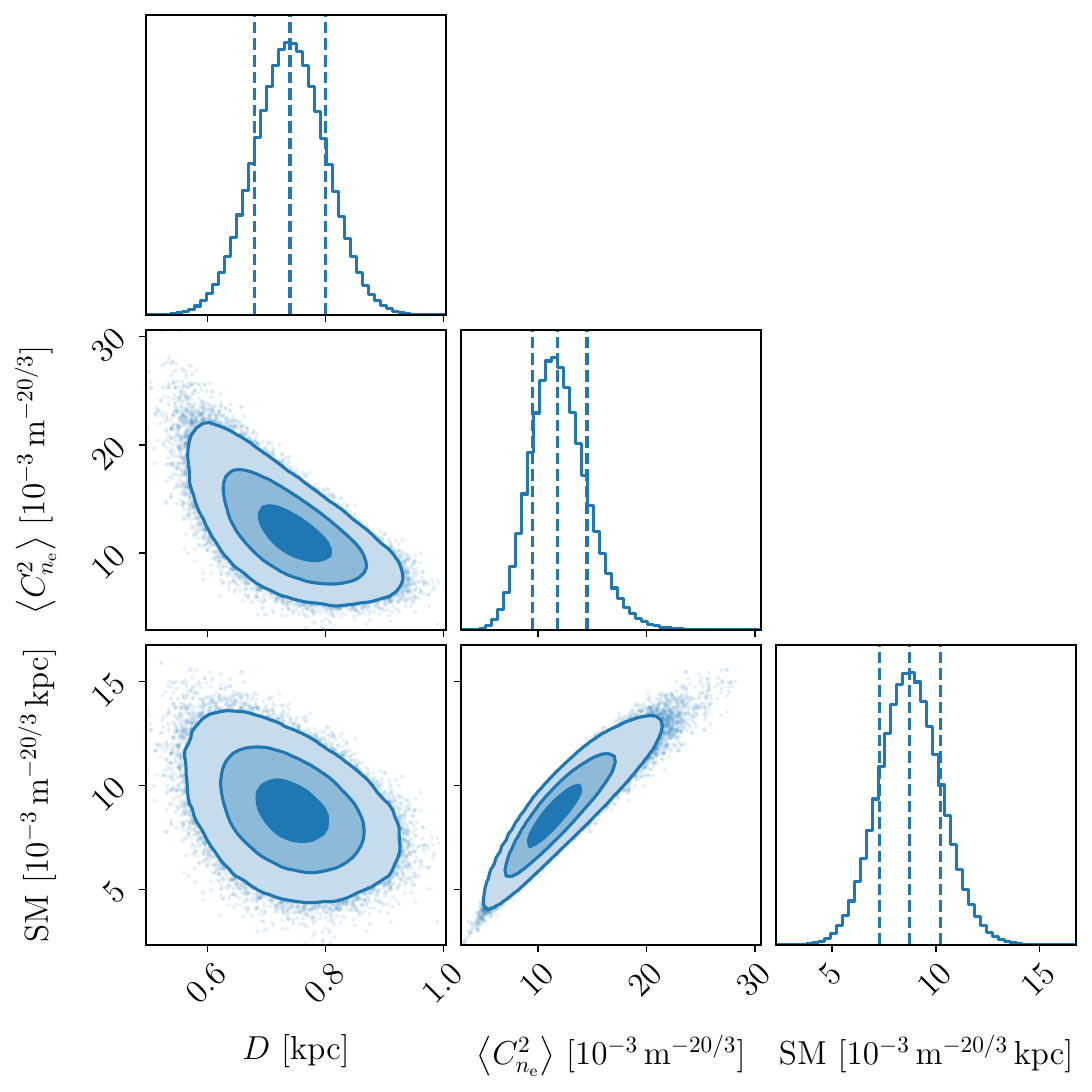}
    \caption{Corner plot showing covariances between the pulsar distance ($D$), average scattering strength along the line of sight ($\left\langle C_{n_\mathrm{e}}^2 \right\rangle$), and scattering measure (SM) for J0737$-$3039A. The 1D distributions show the medians and \qty{68}{\percent} confidence intervals. The 2D distributions show the \qty{39.3}{\percent}, \qty{86.5}{\percent}, and \qty{98.9}{\percent} confidence levels.}
    \label{fig:scat-corner}
\end{figure}

Stochastic fluctuations in the electron density in the IISM ($n_\mathrm{e}$) are characterized by a power-law wavenumber spectrum between two cutoffs known as the ``inner" and ``outer" scale.
This is written as
\begin{equation}
    P_{n_\mathrm{e}}(\kappa) = C_{n_\mathrm{e}}^2 \kappa^{-\beta},\qquad \kappa_\mathrm{o} \leq \kappa \leq \kappa_\mathrm{i},
\end{equation}
where $\kappa$ is the wavenumber, $\kappa_\mathrm{i}$ and $\kappa_\mathrm{o}$ are the wavenumbers of the inner and outer scale, respectively, $C_{n_\mathrm{e}}^2$ is the amplitude of the power spectrum (i.e.\ the scattering strength) with units of $\mathrm{m}^{-20/3}$, and $\beta$ is the spectral index \citep{Rickett1977}.
Since a medium exhibiting Kolmogorov turbulence ($\beta=11/3$) can be described analytically, it is common to use this as a simplifying assumption to obtain approximate values for the key properties of the scattering medium \citep[e.g.][]{Bhat2004}.
For example, the average $C_{n_\mathrm{e}}^2$ along the line of sight, $\left\langle C_{n_\mathrm{e}}^2 \right\rangle$, can be estimated by assuming the turbulence is distributed homogeneously and exhibits Kolmogorov turbulence.
Then,
\begin{multline}
    \left\langle C_{n_\mathrm{e}}^2 \right\rangle \simeq \num{2e-3}\,\mathrm{m}^{-20/3} \\
    \times \bigg(\frac{\nu}{\unit{GHz}}\bigg)^{11/3} \bigg(\frac{D}{\unit{\kpc}}\bigg)^{-11/6} \bigg(\frac{\Delta\nu_\mathrm{d}}{\unit{\MHz}}\bigg)^{-5/6},
\end{multline}
where $D$ is the distance to the pulsar \citep{Cordes1985}.
The scattering measure (SM) is the integrated scattering strength along the line of sight -- a useful quantity for comparing the scattering properties of different lines of sight \citep[e.g.][]{Cordes1991}.
Under the above assumptions, it may be calculated as
\begin{equation}\label{eq:sm}
    \mathrm{SM} = D\int_\mathcal{L} C_{n_\mathrm{e}}^2(s)\,\mathrm{d}s = D\left\langle C_{n_\mathrm{e}}^2 \right\rangle,
\end{equation}
where the integral is performed over the propagation path $\mathcal{L}$ parameterized by the fractional distance $s$ that ranges from $s=0$ at the pulsar to $s=1$ at the observer.

We performed this analysis using our best estimates of the pulse broadening at \qty{150}{\MHz}.
For J0737$-$3039A and J0742$-$2822, we used the joint spectral fits with the literature based on the Kolmogorov assumption (Table~\ref{tab:alpha-lit}).
For J0837$-$4135, we used our in-band spectral fit (Table~\ref{tab:alpha}).
The $\tau_\mathrm{s}$ estimates were converted into equivalent $\Delta\nu_\mathrm{d}$ using Equation~\eqref{eq:scat-scin-relation} assuming $C_1=0.654$ for Kolmogorov turbulence \citep{Lambert1999}.

To estimate the uncertainties of $\Delta\nu_\mathrm{d}$, $\left\langle C_{n_\mathrm{e}}^2 \right\rangle$, and SM, we simulated their distributions by drawing a large number of random samples from assumed distributions of $\tau_\mathrm{s}$ and $D$.
For J0742$-$2822 and J0837$-$4135, the broad and asymmetric uncertainties in $D$ (cf.\ Table~\ref{tab:pulsars}) were modeled as log-normal distributions with upper limits from \citet{Verbiest2012} so that they were well-behaved when propagating through the equations.
The parameter estimates are summarized in Table~\ref{tab:sm} and a corner plot illustrating the result for J0737$-$3039A is shown in Figure~\ref{fig:scat-corner}.
See Section \ref{sec:gum-nebula} for further discussion on the SM estimates.

\subsection{Probing the Gum Nebula}\label{sec:gum-nebula}

\begin{deluxetable}{lccccccccc}
\tablewidth{0pt}
\tablecaption{Log-differences between our nominal $\tau_\mathrm{s}$ and SM measurements and the predictions from the NE2001 and NE2025 models.\label{tab:ne20xx-compare}}
\tablehead{
    PSR & \colhead{Clump(s)} &
    \multicolumn{4}{c}{$\log_{10}\big(\tau_\mathrm{s,mod}/\tau_\mathrm{s,obs}\big)$ at \qty{1}{\GHz}} &
    \multicolumn{4}{c}{$\log_{10}\big(\mathrm{SM}_\mathrm{mod}/\mathrm{SM}_\mathrm{obs}\big)$} \\[2pt]
    \cline{3-6}\cline{7-10}
    & \colhead{intersected} &
    \multicolumn{2}{c}{NE2001} & \multicolumn{2}{c}{NE2025} &
    \multicolumn{2}{c}{NE2001} & \multicolumn{2}{c}{NE2025} \\
    \cline{3-4}\cline{5-6}\cline{7-8}\cline{9-10}
    & &
    \colhead{DM} & \colhead{$D$} & \colhead{DM} & \colhead{$D$} &
    \colhead{DM} & \colhead{$D$} & \colhead{DM} & \colhead{$D$}
}
\startdata
J0737$-$3039A & Gum I & \num{-3.7} & \num{-3.4} & 0.67 & 1.5 & \num{-3.6} & \num{-3.4} & 0.41 & 0.54 \\
J0742$-$2822  & Gum I & \num{-1.3} & \num{-1.6} & 1.4  & 1.4 & \num{-2.0} & \num{-2.2} & \num{-0.018} & \num{-0.02} \\
J0837$-$4135  & Gum I \& II, Vela & 0.75 & 0.86 & 1.4 & 1.9 & 0.15 & 0.15 & 1.0 & 1.1 \\
\enddata
\tablecomments{The DM and $D$ columns correspond to the DM-derived and distance-derived model predictions, respectively.}
\end{deluxetable}

\begin{figure}
    \centering
    \includegraphics[width=\linewidth]{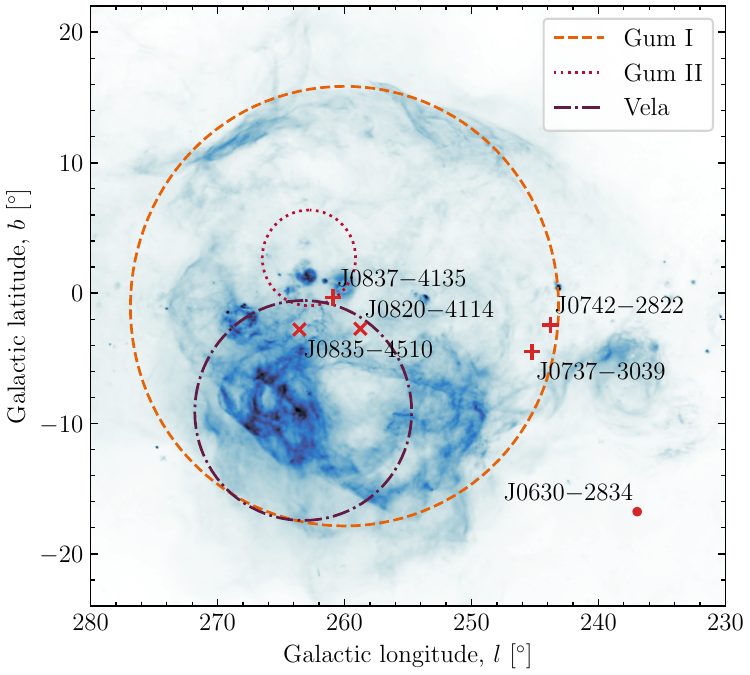}
    \caption{Sky locations of the target pulsars in Galactic coordinates. The background image is a H$\alpha$ intensity map on a linear color scale showing the Gum Nebula \citep{Finkbeiner2003}. The markers indicate whether the pulsars are within ($\times$), behind (+), or outside of (filled circle) the Gum Nebula. The circular regions indicate the projected angular sizes of the Gum I, Gum II, and Vela clumps in the NE2025 model \citep{NE2025}.}
    \label{fig:hmap}
\end{figure}

The Gum Nebula is a large shell of ionized material in the IISM that contributes significantly to the scattering and dispersion of pulsars within and behind it \citep{Gum1952}.
As illustrated on the sky map in Figure~\ref{fig:hmap}, this includes all of our target pulsars except J0630$-$2834.
The Gum Nebula region is therefore a significant component in models of the Galactic electron density distribution such as TC93 \citep{Taylor1993}, NE2001 \citep{NE2001}, and NE2025 \citep{NE2025}.
The scattering strength is defined in these models as 
\begin{equation}\label{eq:cn-ne2001}
    C_{n_\mathrm{e}}^2(s)=C_\mathrm{SM}F_\mathrm{c}(s)n_\mathrm{e}^2(s),
\end{equation}
where $C_\mathrm{SM}$ is a constant and $F_\mathrm{c}$ is the density fluctuation parameter \citep{Taylor1993}.
The SM is then calculated from the integral defined in Equation~\eqref{eq:sm}.
The pulse broadening timescale is calculated from the SM under the assumption of Kolmogorov turbulence in a uniform medium:
\begin{equation}\label{eq:tau-from-sm}
    \tau_\mathrm{s} \simeq \qty{1.10}{\ms} \bigg(\frac{\mathrm{SM}}{\mathrm{m}^{-20/3}\,\unit{\kpc}}\bigg)^{6/5} \bigg(\frac{\nu}{\unit{\GHz}}\bigg)^{-22/5} \bigg(\frac{D}{\unit{\kpc}}\bigg)
\end{equation}
\citep{NE2001}.

In NE2001 and NE2025, the Gum Nebula region is separated into three spherical Gaussian clumps known as Gum I, Gum II, and Vela (shown in Figure~\ref{fig:hmap}).
It was noted by \citet{NE2025} that $\tau_\mathrm{s}$ was underestimated by NE2001 for pulsars viewed through Gum I.
For NE2025, the fluctuation parameters of the three clumps were jointly fit using scattering measurements from 17 pulsars, resulting in around an order of magnitude improvement in the predicted $\tau_\mathrm{s}$.
Specifically, they minimized the mean and standard deviation of $\log_{10}\big(\tau_\mathrm{s,mod}/\tau_\mathrm{s,obs}\big)$, where $\tau_\mathrm{s,mod}$ and $\tau_\mathrm{s,obs}$ are the model prediction and the observed value of the pulse broadening timescale at \qty{1}{\GHz}.
They reported a minimized mean of \num{-0.0003} and standard deviation of \num{0.59}.
As we have obtained new and precise scattering measurements for three pulsars behind the Gum Nebula, it is of interest to compare our measurements to the model predictions.
In Table~\ref{tab:ne20xx-compare} we summarize the log-differences between the predicted and observed $\tau_\mathrm{s}$ and SM for NE2001 and NE2025.
We discuss these differences for each pulsar below, with the intention of informing future refinements to the NE2025 model.

\subsubsection{PSR J0737\safeminus{}3039A}\label{sec:necomp:0737}

Our strongest constraint comes from J0737$-$3039A due to its precisely known distance of $D=0.74(6)\,\unit{\kpc}$ obtained from parallax measurements \citep{Kramer2021PhRvX}.
\citet{Askew2026} used scintillation measurements to independently verify the parallax distance and estimate that the scattering screen is located at a distance of $0.36^{+0.03}_{-0.04}\,\unit{\kpc}$, making it likely associated with the edge of the Gum Nebula (see Figure~\ref{fig:hmap}).
In NE2001 and NE2025, the electron density contribution for this line of sight comes almost entirely from Gum I and the DM is overestimated by \qty{16.3}{\per\cm\cubed\pc} and \qty{18.0}{\per\cm\cubed\pc}, respectively.
The fluctuation parameter for Gum I was first assigned a non-zero value in NE2025.
As a result, the SM and $\tau_\mathrm{s}$, which are both underestimated by 3.4 orders of magnitude by NE2001, are overestimated by 1.5 and 0.5 orders of magnitude, respectively, by NE2025.\footnote{The differences in the size of the corrections comes from the relationship $\tau_\mathrm{s}\propto\mathrm{SM}^{6/5}$ in Equation~\eqref{eq:tau-from-sm}.}.
The overestimation of the DM, SM, and $\tau_\mathrm{s}$ point to an overestimation of the integrated electron content of Gum I.
This could either be due to an overestimation of the electron density or the size of the clump.
However, these observations do not rule out a simultaneous overestimation of $F_\mathrm{c}$, which could contribute to the overestimated scattering parameters.

\subsubsection{PSR J0742\safeminus{}2822}

The distance to J0742$-$2822 was estimated by \citet{Verbiest2012} to be $D=2.0^{+1.0}_{-0.8}\,\unit{\kpc}$ using H\textsc{i} absorption line constraints from \citet{Koribalski1995}.
This places the pulsar far beyond the Gum Nebula.
\citet{Wang2025} used scintillation measurements to estimate that the scattering screen is located at $0.25(7)\,\unit{\kpc}$, significantly closer than that of J0737$-$3039A.
Interestingly, although the line of sight intersects a similar point in Gum I to J0737$-$3039A (see Figure~\ref{fig:hmap}), NE2001 and NE2025 underestimate the DM by \qty{4.4}{\per\cm\cubed\pc} and \qty{9.3}{\per\cm\cubed\pc}, respectively.
As with J0737$-$3039A, NE2001 significantly underestimates $\tau_\mathrm{s}$ and SM for J0742$-$2822.
In NE2025, the SM is in close agreement with our measurement, but $\tau_\mathrm{s}$ is overestimated by 1.4 orders of magnitude.
Since the DM is slightly underestimated, but the SM is not, this points to a possible overestimation of $F_\mathrm{c}$.
Additionally, since the SM is consistent with our measurement and $\tau_\mathrm{s}$ is significantly overestimated, this points to an overestimation of the pulsar distance.

\subsubsection{PSR J0837\safeminus{}4135}

This pulsar lies near the intersection of Gum I, Gum II, and Vela (see Figure~\ref{fig:hmap}) at an estimated distance of $D=1.5^{+1.2}_{-0.9}\,\unit{\kpc}$ based on H\textsc{i} measurements \citet{Verbiest2012,Johnston1996}.
It therefore provides a particularly valuable line of sight for constraining the model parameters.
In NE2001, the line of sight intersects Gum I and Gum II (but not Vela) and the DM is relatively accurate, with an overestimation of \qty{19.3}{\per\cm\cubed\pc} above its observed DM of \qty{147.3}{\per\cm\cubed\pc}.
The predicted SM and $\tau_\mathrm{s}$ are both within an order of magnitude of our measurement, indicating that the Gum II clump is likely the dominant scattering medium.
In NE2025, the Vela clump was moved further away and is radius was increased proportionally to maintain the same angular size.
As a result, the line of sight to J0837$-$4135 intersects the Vela clump, causing the DM to be overestimated by \qty{112.0}{\per\cm\cubed\pc}.
The SM and $\tau_\mathrm{s}$ are also overestimated by an order of magnitude in NE2025 compared with NE2001.
Given that our measurements are more consistent with the model predictions when the Vela clump does not intersect the line of sight, this may suggest that a reduction of its radius could improve the model.
Another possibility is that the spherically-symmetric Gaussian electron distribution is too simplistic to account for the scattering of all the pulsars in the vicinity of Vela.

\subsubsection{Other pulsars}

\paragraph{PSR J0835\safeminus{}4510}
The Vela pulsar is a unique case as it resides within the highly-turbulent environment of the Vela supernova remnant (SNR).
\citet{Johnston1998} measured the scattering spectral index to be \num{-3.93} over a factor of 100 in frequency using a combination of pulse broadening and scintillation measurements, noting that this is consistent with scattering by discrete sources in the Vela SNR or the Gum Nebula.
This is supported by long-term variations in the DM and Faraday rotation measure (RM), which indicate the presence of a discrete magnetized structure along the line of sight that may be a filament in the Vela SNR \citep{Hamilton1977,Hamilton1985,Lee2024}.

\paragraph{PSR J0820\safeminus{}4114}
Based on the DM of this pulsar (\qty{113.34}{\per\cm\cubed\pc}), the NE2025 model predicts $\tau_\mathrm{s}=\qty{71.2}{\us}$ at \qty{1}{\GHz}, which extrapolates to $\sim$\qty{140}{\ms} ($\mathbin{\sim}0.25P$) at \qty{150}{\MHz} assuming $\alpha=-4$.
Therefore, although it is difficult to quantify the constraints on $\tau_\mathrm{s}$, it is clear that this pulsar is significantly less scattered than expected.
If the pulsar is located at the DM-derived distance of \qty{320}{\pc}, then it may be close to the nearside edge of the Gum Nebula.
In that case, the unexpectedly weak scattering could be due to the screen being located relatively close to the pulsar.

\subsection{Profile evolution of PSR J0737\safeminus{}3039A}

The average pulse profile of J0737$-$3039A consists of two widely-separated pulses that correspond to beams with a similar radius from opposite magnetic poles \citep{Guillemot2013,Kramer2021MNRAS}.
As noted in \citet{Kramer2008}, the leading (main) pulse shows significant evolution with frequency, whereas the trailing (inter)-pulse shows minimal frequency evolution (cf.\ their Figure~3 and our Figure~\ref{fig:J0737-profiles}).
Our \qty[number-unit-product=-]{227}{\MHz} profile is consistent with the \qty[number-unit-product=-]{325}{\MHz} profile published in \citet{Kramer2008}.
In the top three MWA subbands, we see a continuation of the trend observed at higher frequencies, where the core component increases in amplitude relative to the rest of the profile with decreasing frequency.
In the lowest subband, at \qty{113}{\MHz}, the trailing pulse is no longer detectable.

At higher frequencies (above \qty{600}{\MHz}), it is particularly evident that the main pulse can be roughly described as a ``core" component flanked by two ``cone" components \citep[a morphological distinction common for non-recycled pulsars; e.g.][]{Rankin1983}.
The core component has a steeper spectral index than the rest of the profile, and also appears to be associated with a different orthogonal polarization mode \citep[see Figure~8 of][]{Kramer2021MNRAS}.
This behavior could potentially be explained by birefringence in the pulsar magnetosphere, which causes the ordinary (O) mode to be refracted along the magnetic field lines while the extraordinary (X) mode propagates along its initial trajectory without being refracted \citep{Barnard1986}.
This can, for example, lead to broadening of the profile as the modes separate at low frequencies \citep[e.g.][]{McKinnon1997}.
In the case of J0737$-$3039A, the profile evolution could be explained by the core component being dominated by X-mode emission and the rest of the profile being dominated by O-mode emission.
As such, the interpulse and cone components would be refracted out of the line of sight at lower frequencies, leading to a steeper flux density spectrum for these components.

\subsection{Prospects for timing PSR J0737\safeminus{}3039A at low frequencies}

One of the aims of this work is to investigate the timing potential of J0737$-$3039A below \qty{300}{\MHz}, and the viability of DM monitoring to improve long-term timing results.
Future pulsar observations with the MWA will make use of the real-time beamforming capability and increased number of available tiles in Phase III \citep{Tingay2026}.
For timing observations, we consider only frequencies for which the TOA precision is not significantly affected by scattering.
Based on the results from Section~\ref{sec:res:dm}, we know that frequencies of $\sim$\qty{176}{\MHz} and higher (up to \qty{300}{\MHz}) are usable.
In our observations at these frequencies, we found TOA uncertainties of $\sigma_\mathrm{TOA}\sim\qty{50}{\us}$.
Ideally, $\sigma_\mathrm{TOA}$ should scale proportionally with S/N \citep[e.g.][]{LorimerKramer2012}.
Therefore, for the same integration time (\qty{900}{\s}) and bandwidth (\qty{3.24}{\MHz}), we may expect the following scaling for MWA observations:
\begin{equation}
    \sigma_\mathrm{TOA} \simeq \qty{50}{\us} \bigg(\frac{N_\mathrm{T}}{110}\bigg)^{-1},
\end{equation}
where $N_\mathrm{T}$ is the number of tiles.
For the full array with $N_\mathrm{T}=256$, this scaling predicts $\sigma_\mathrm{TOA}\sim\qty{20}{\us}$.
To first order, the DM uncertainty $\sigma_\mathrm{DM}$ scales with $\sigma_\mathrm{TOA}$, the number of TOAs, $N_\mathrm{TOA}$, and the fractional frequency span of the observations, $\Delta\tilde{\nu}$, as follows:
\begin{multline}
    \sigma_\mathrm{DM} \simeq \qty{4e-4}{\per\cm\cubed\pc} \\
    \times \bigg(\frac{\sigma_\mathrm{TOA}}{\qty{50}{\us}}\bigg) \bigg(\frac{N_\mathrm{TOA}}{40}\bigg)^{-1/2} \bigg(\frac{\Delta\tilde{\nu}}{0.27}\bigg)^{-1}.
\end{multline}
Thus, if we redistribute all four subbands between \qtyrange{176}{300}{\MHz} (so that $N_\mathrm{TOA}=80$ and $\Delta\tilde{\nu}\approx0.52$) then for the same number of tiles we may expect $\sigma_\mathrm{DM}\sim\qty{1.5e-4}{\per\cm\cubed\pc}$.
For $N_\mathrm{T}=256$, this becomes $\sigma_\mathrm{DM}\sim\qty{6e-5}{\per\cm\cubed\pc}$.

There are several caveats to the scaling relations described above.
First, the tied-array sensitivity does not scale linearly with the number of tiles, particularly for the extended baselines which de-phase more rapidly and, therefore, gradually contribute more noise to the coherent sum of tile signals across the array.
For this reason, the 200-tile ``Phase I plus Hexes" configuration \citep{Tingay2026} may be optimal for tied-array sensitivity.
In this configuration, we expect $\sigma_\mathrm{DM}\sim\qty{8e-5}{\per\cm\cubed\pc}$.
Second, we should expect a net loss of sensitivity at higher frequencies due to the intrinsic flux-density spectrum of the pulsar \citep[see the discussion in][]{Lee2026}.
The flux-density spectrum below \qty{300}{\MHz} is not well constrained, with conflicting measurements from different publications \citep[cf.\ the \textsc{pulsar\_spectra} flux density catalog;][]{Swainston2022}.
Even between the observations from this work, we see differences in S/N that are likely due to a combination of calibration quality and RISS.
As such, to accurately characterize the spectral dependence of the S/N, it is desirable to simultaneously observe the pulsar between $\sim$\qtyrange{200}{300}{\MHz}.

In the 16-year timing study of J0737$-$3039A reported by \citet{Kramer2021PhRvX}, the DM precision was of the order $\numrange{5}{10}\times10^{-4}\,\unit{\per\cm\cubed\pc}$ when averaged into 100-d bins.
Thus, even from single observing epochs between \qtyrange{176}{231}{\MHz} using half of the total available recording bandwidth of the MWA, we are able to obtain a precision comparable to the best precision of the 100-d average from previous work.
With the full capabilities of the MWA Phase III, we expect the MWA to improve upon the precision of past DM monitoring by around an order of magnitude when averaged over the same time span.
The SKA-Low will enable another large improvement in DM precision.
As such, regular and long-term monitoring with the MWA and SKA-Low can, in principle, significantly aid in the improvement of timing precision for the double pulsar.

%% file: 5-Conclusions.tex
\section{Conclusions and Future Work}\label{sec:dp:conclusions}

We have reported on the first in-depth study of PSR J0737$-$3039A below \qty{300}{\MHz} using four dedicated full-orbit observations with the MWA-VCS.
The \qty[number-unit-product=-]{30.72}{\MHz} recording bandwidth was distributed between four equal subbands spanning \qtyrange{109}{231}{\MHz}.
Taking advantage of the large field of view of the MWA and the flexibility of the VCS data, we obtained simultaneous detections of J0737$-$3039A and five nearby non-recycled pulsars in these observations.

Our timing analysis of J0737$-$3039A yielded a DM precision of \qty{4e-4}{\per\cm\cubed\pc} using only half of the available bandwidth of the MWA over the frequency range \qtyrange{176}{231}{\MHz}.
We estimate that the MWA Phase III is capable of obtaining a precision of \qty{8e-5}{\per\cm\cubed\pc} using the 200-tile configuration.
The new real-time beamforming capability of the Phase III system will enable regular monitoring at low frequencies for the first time.
In principle, the MWA could provide around an order of magnitude improvement in constraints on temporal DM variations compared with previous long-term monitoring at higher frequencies.

For the first time, we have measured pulse broadening for J0737$-$3039A, which we find becomes significant (i.e.\ exceeds the intrinsic pulse width) below $\sim$\qty{180}{\MHz}.
Within the MWA band, we measure the scattering spectral index to be $\alpha=\num{-3.1(4)}$.
When combining our data with scintillation bandwidths from MeerKAT, together spanning \qtyrange{0.1}{2.6}{\GHz}, we find that $\alpha$ is highly correlated with the proportionality constant $C_1$ (a proxy for the spectral and geometric properties of the scattering medium).
Assuming a thin-screen geometry, the data are more consistent with a square-law wavenumber spectrum than a Kolmogorov wavenumber spectrum.
For a square-law wavenumber spectrum, we measure $\alpha=\num{-4.1(1)}$.
We have also placed new and improved constraints on the scattering spectra of PSRs J0742$-$2822 and J0837$-$4135.
For J0742$-$2822, we combined our measurements with scintillation bandwidths from the TMRT, together spanning \qtyrange{0.1}{9.0}{\GHz}.
Again, a square-law wavenumber spectrum is preferred, which results in $\alpha=\num{-3.8(1)}$.
For J0837$-$4135, we measure $\alpha=\num{-4.1(1)}$ within the MWA band and found that the data are in good agreement with a power-law spectrum.
Thus, we observe a deviation from Kolmogorov turbulence for all three pulsars.
This may be due to scattering by discrete structures within the Gum Nebula region.

We compared our scattering measurements with the NE2025 model of the Galactic electron density distribution.
Assuming the pulsar distances are accurate, our measurements for J0737$-$3039A and J0742$-$2822, which have similar lines of sight through the Gum Nebula, point to a possible overestimation of the fluctuation parameter $F_\mathrm{c}$ of the Gum I clump.
Additionally, the DM is overestimated by the model towards J0737$-$3039A, but not J0742$-$2822, indicating that the electron distribution model of Gum I may be too simplistic in this region.
Our measurements also suggest that the Vela clump causes a significant overestimation of the DM and scattering for J0837$-$4135.
This may be explained by an overestimation of the size of the Vela clump.

Through this work, we have demonstrated the unique capability of the MWA to simultaneously probe scattering for multiple lines of sight throughout the Gum Nebula region.
The pulse broadening for all of these pulsars, besides the Vela pulsar, is either minimal or undetectable above \qty{300}{\MHz}.
This places the MWA in a unique position to characterize their scattering at low frequencies.
In future, similar studies of the Gum Nebula region with the MWA Phase III could target more pulsars with even greater sensitivity.
The real-time beamformer will also make it practical to collect regular observations over several refractive cycles in order to minimize systematic bias due to scintillation.
This will provide valuable improvements to pulse broadening timescale estimates, which can be used to inform future Galactic electron density models.

%% file: 6-Acknowledgements.tex
\begin{acknowledgements}
We thank Chia Min Tan and Ryan Craft for useful discussions, and the anonymous reviewer for constructive feedback.
C.P.L. was supported by an Australian Government Research Training Program (RTP) Stipend and RTP Fee-Offset Scholarship (\url{https://doi.org/10.82133/C42F-K220}).
This scientific work made use of data obtained from Inyarrimanha Ilgari Bundara, the Murchison Radio-astronomy Observatory, operated by CSIRO.
We acknowledge the Wajarri Yamatji people as the traditional owners of the Observatory site.
This work was supported by resources provided by the Pawsey Supercomputing Research Centre’s Setonix Supercomputer (\url{https://doi.org/10.48569/18sb-8s43}), and their Acacia (\url{https://doi.org/10.48569/nfe9-a426}) and Banksia (\url{https://doi.org/10.48569/tnja-4s30}) Object Storage systems, with funding from the Australian Government and the Government of Western Australia.
This work made use of NASA's Astrophysics Data System and arXiv.
\end{acknowledgements}

\facilities{MWA(VCS)}

\software{
\textsc{mwa\_hyperdive} \citep{hyperdrive},
\textsc{vcsbeam} \citep{Bhat2026},
\textsc{dspsr} \citep{vanStraten2011},
\textsc{psrchive} \citep{Hotan2004,vanStraten2012},
\textsc{tempo2} \citep{Tempo2},
\textsc{clfd} \citep{Morello2019,clfd},
\textsc{matplotlib} \citep{Matplotlib},
\textsc{numpy} \citep{NumPy},
\textsc{scipy} \citep{2020SciPy-NMeth},
\textsc{astropy} \citep{Astropy2013,Astropy2018,Astropy2022}.
}

%% file: 7-Appendix.tex
\section{Observations Summary}\label{app:observations}

Each observation was performed in three contiguous \qty[number-unit-product=-]{50}{\minute} subobservations in order keep the double pulsar near the tile pointing center.
In Table~\ref{tab:subobs-summary}, we provide information about each subobservation, including the MWA observation ID, the UTC start time, the number of ``good" tiles, and the tile pointing center.

\begin{deluxetable}{lccccc}[h]
\tablewidth{0pt}
\tablecaption{Summary of subobservations. \label{tab:subobs-summary}}
\tablehead{
    Label &
    Part &
    Obs ID &
    \colhead{Start Time} &
    \colhead{$N_\mathrm{tile}$} &
    \colhead{(Az, El)} \\[-5pt]
    &
    &
    &
    \colhead{(UTC)} &
    &
    \colhead{[\unit{\degree},\unit{\degree}]}
}
\startdata
Epoch A & 1 & 1387125264 & 2023-12-20 16:34:06 & 104 & (116.6, 74.6) \\
        & 2 & 1387128264 & 2023-12-20 17:24:06 & 105 & (180.0, 83.2) \\
        & 3 & 1387131264 & 2023-12-20 18:14:06 & 104 & (225.0, 80.3) \\
Epoch B & 1 & 1389796344 & 2024-01-20 14:32:06 & 118 & (116.6, 74.6) \\
        & 2 & 1389799344 & 2024-01-20 15:22:06 & 115 & (180.0, 83.2) \\
        & 3 & 1389802344 & 2024-01-20 16:12:06 & 118 & (225.0, 80.3) \\
Epoch C & 1 & 1420645352 & 2025-01-11 15:42:14 &  85 & (135.0, 80.3) \\
        & 2 & 1420648352 & 2025-01-11 16:32:14 &  85 & (225.0, 80.3) \\
        & 3 & 1420651352 & 2025-01-11 17:22:14 &  85 & (243.4, 74.6) \\
Epoch D & 1 & 1424433552 & 2025-02-24 11:58:54 & 111 & (116.6, 74.6) \\
        & 2 & 1424436552 & 2025-02-24 12:48:54 & 109 & (135.0, 80.3) \\
        & 3 & 1424439552 & 2025-02-24 13:38:54 & 111 & (225.0, 80.3) \\
\enddata
\end{deluxetable}

\section{Pulse Profiles}\label{app:profiles}

The subband-resolved integrated pulse profiles for each pulsar are provided in Figure~\ref{fig:all-profs}.
It is important to note that the effective on-source time can vary significantly with frequency if the pulsar is located near the edge of the primary beam.
This applies to PSRs J0630$-$2834, J0820$-$4114, J0835$-$4510, and J0837$-$4135, which are located at large angles ($>\qty{10}{\degree}$) from the tile pointing center.
To increase the total integrated S/N of each subband, we flagged the observation times where the pulsar signal is weak or undetectable.
We first formed \qty[number-unit-product=-]{5}{\minute} subintegrations and measured the peak S/N of each profile, $\mathrm{S}/\mathrm{N}_n$, where $n$ is the subintegration index.
The $n$-th subintegration was flagged if $\mathrm{S}/\mathrm{N}_n < 0.4\, Q_{0.95}\left(\mathrm{S}/\mathrm{N}\right)$, where $Q_{0.95}\left(\mathrm{S}/\mathrm{N}\right)$ denotes the 95th percentile of peak S/N measurements for that subband.
For the observations where flagging was applied, the effective integration times are annotated in Figure~\ref{fig:all-profs}.
We found that this method was most effective for brighter detections.
For example, it fails to identify the on-source times in Epoch C for J0630$-$2834 and J0837$-$4135.
For this reason, we have not applied any flagging to J0820$-$4114.

\begin{figure*}[p]
    \centering
    \includegraphics[width=\linewidth]{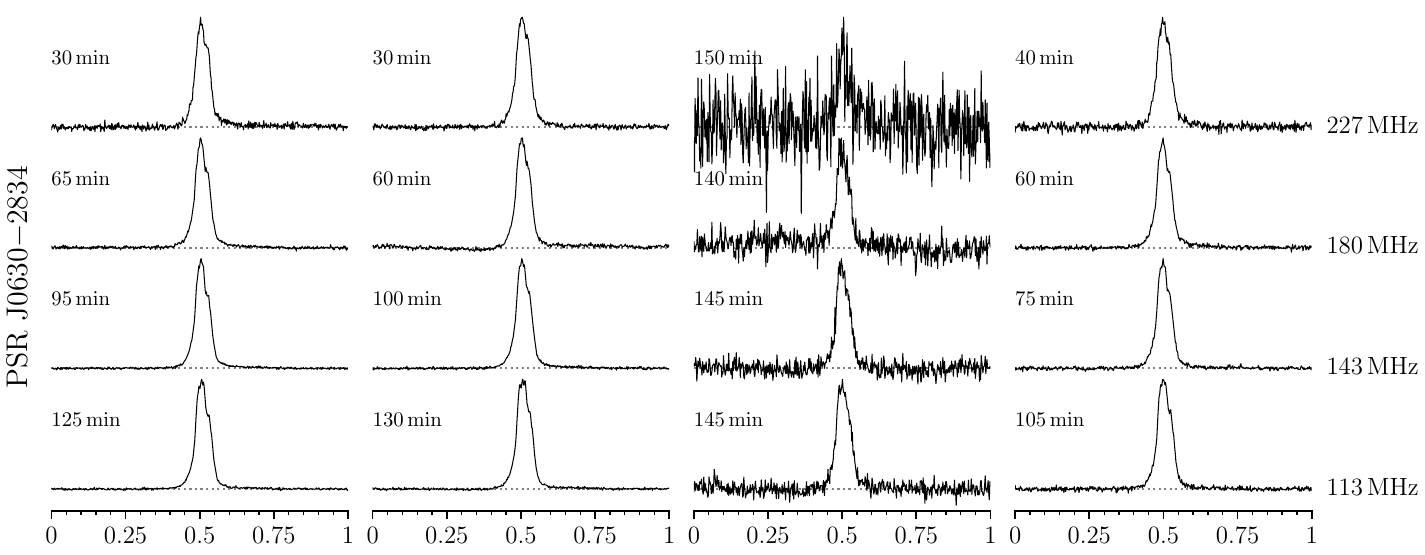}
    \includegraphics[width=\linewidth]{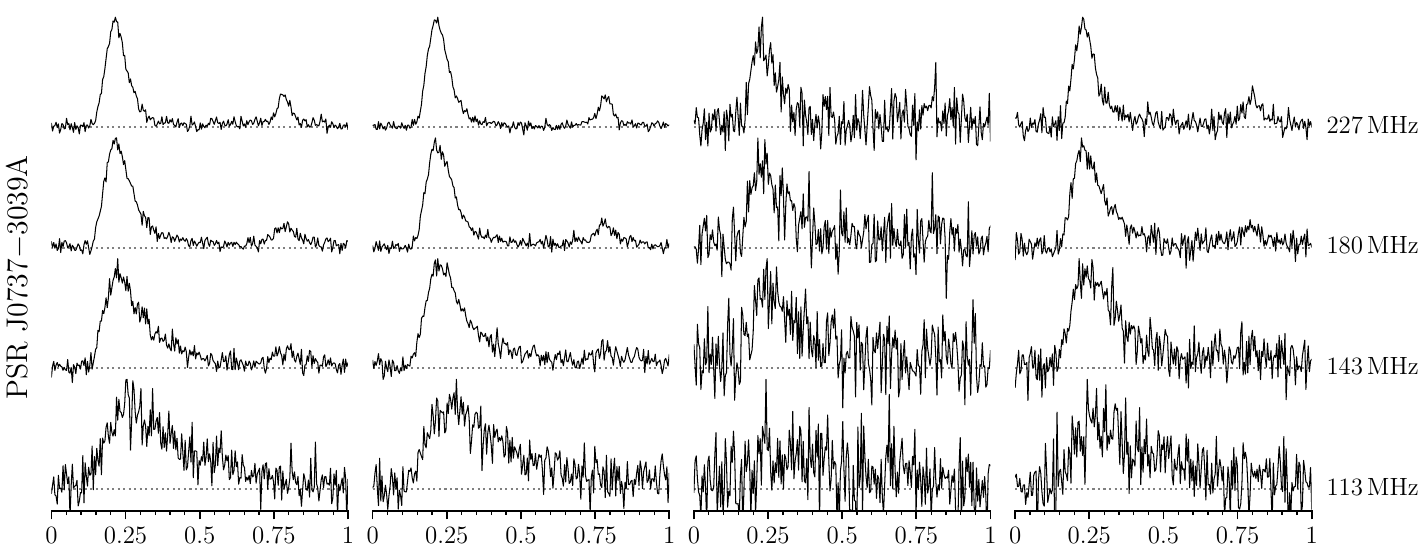}
    \includegraphics[width=\linewidth]{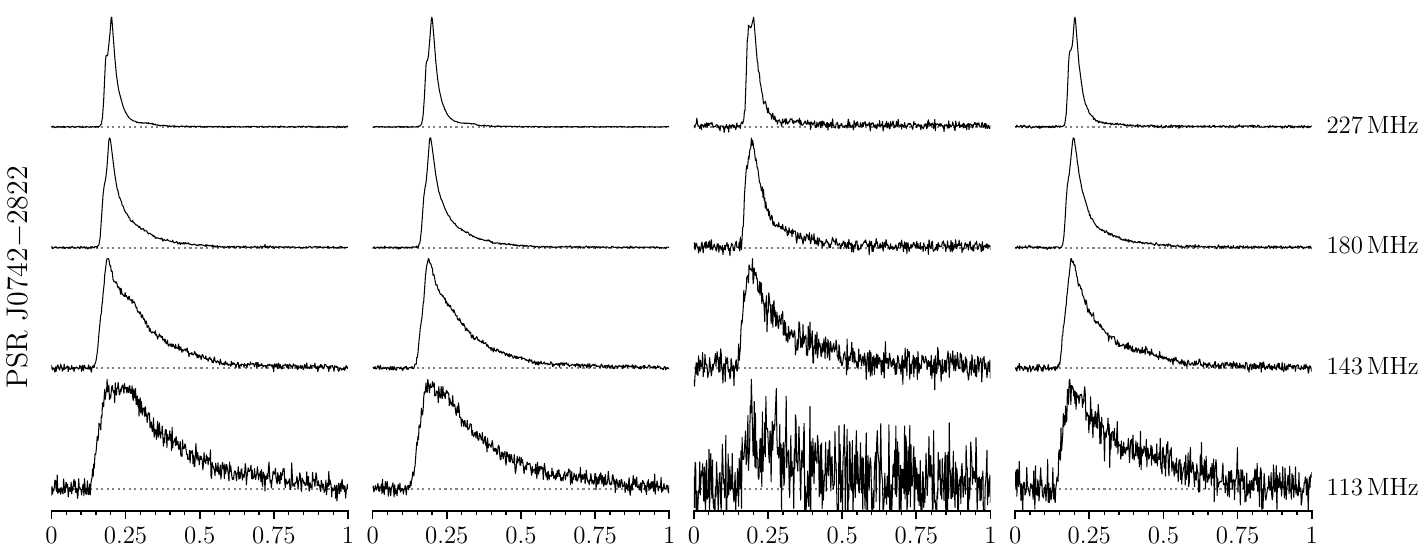}
    \caption{Integrated pulse profiles for the target pulsars (see Table~\ref{tab:pulsars}). From left to right, the columns correspond to detections from Epochs A--D (see Table~\ref{tab:obs-summary}). Each profile has been normalized by the peak intensity. PSR J0737$-$3039A is shown with 256 bins and the other pulsars are shown with 512 bins. The integration times are \qty{150}{\minute} unless otherwise noted.}
    \label{fig:all-profs}
\end{figure*}

\begin{figure*}[p]
    \ContinuedFloat
    \centering
    \includegraphics[width=\linewidth]{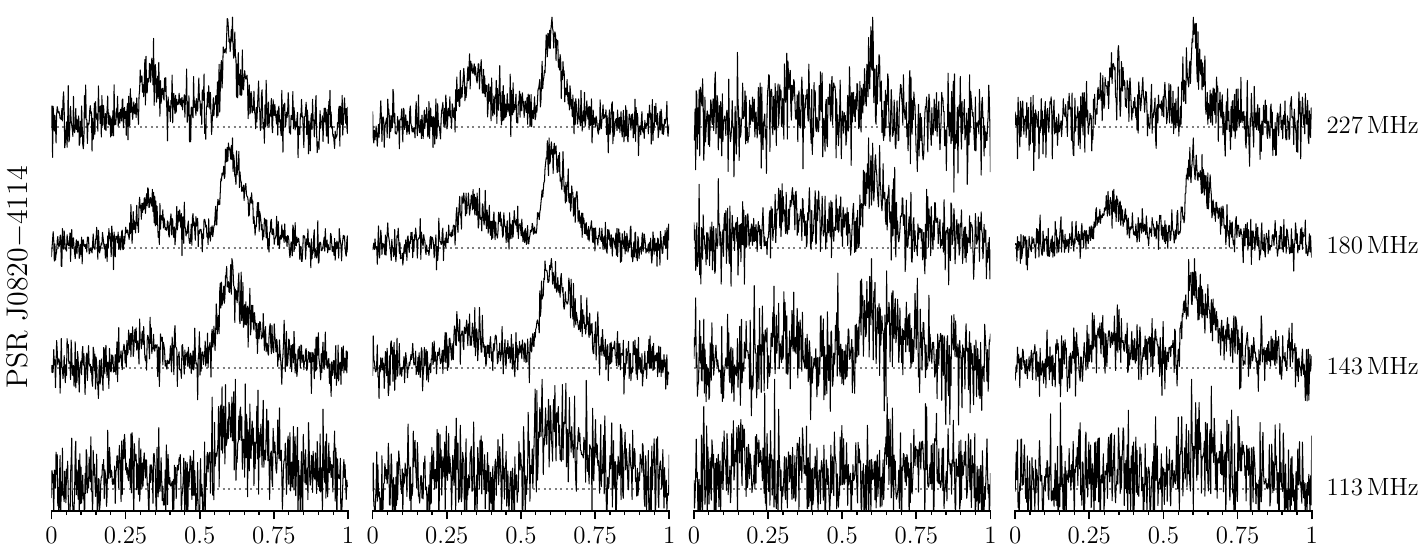}
    \includegraphics[width=\linewidth]{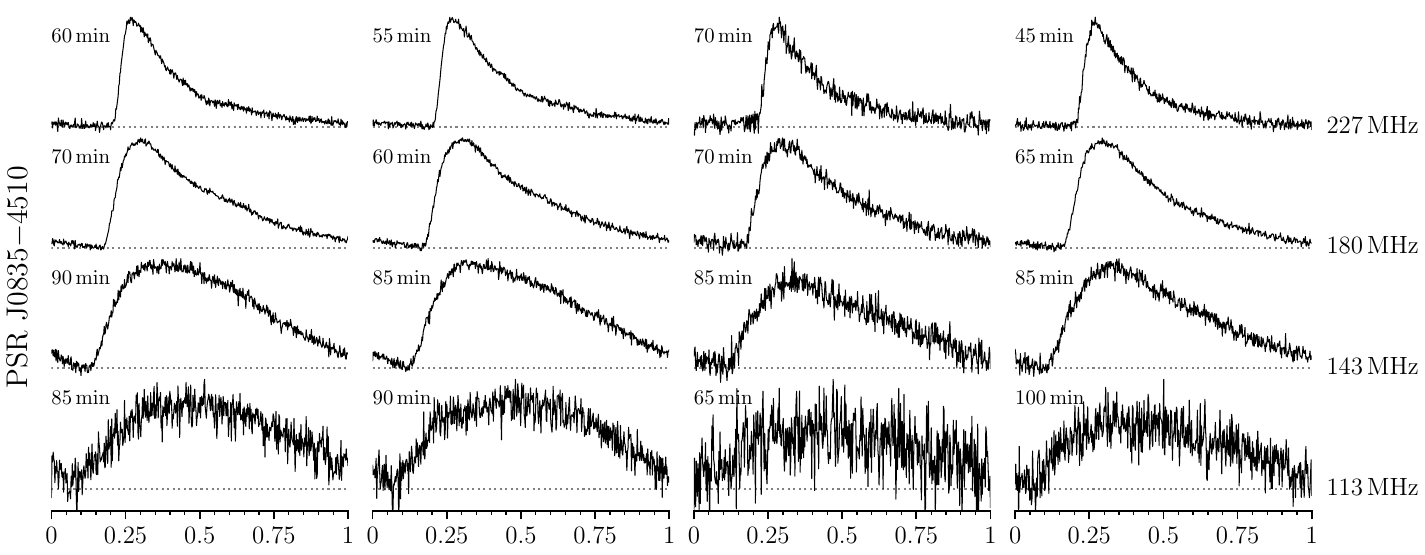}
    \includegraphics[width=\linewidth]{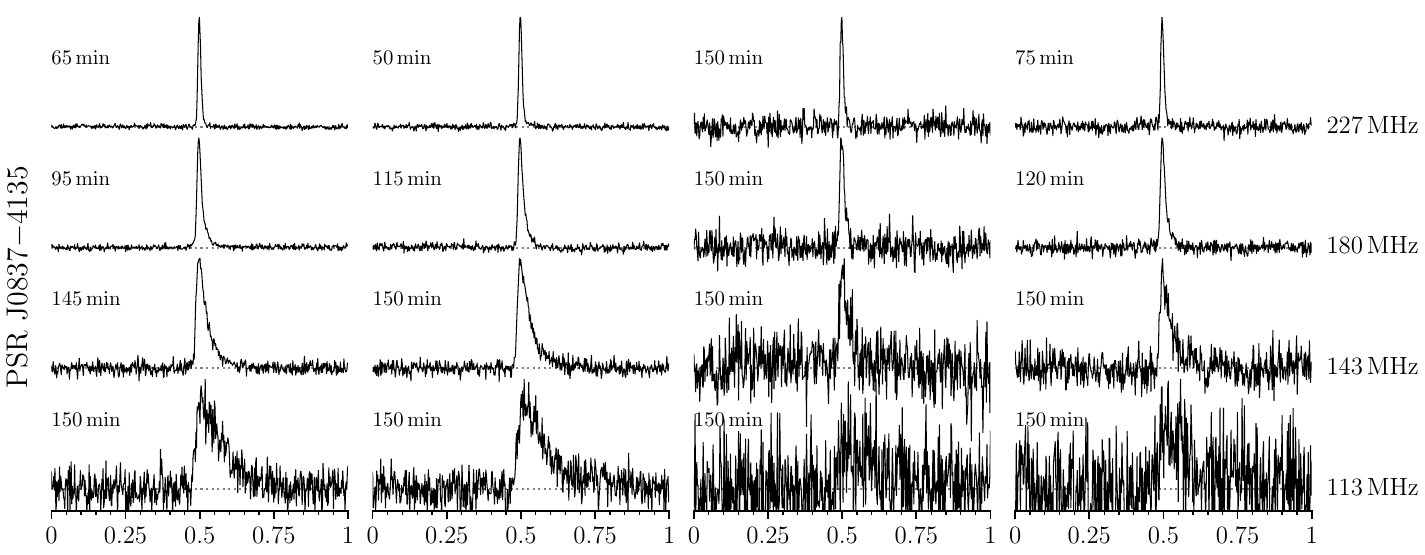}
    \caption{Continued.}
\end{figure*}